\documentclass[12pt]{article}

\usepackage{amssymb,amsmath,amsfonts,amssymb}
\usepackage{graphics,graphicx,color,verbatim}

\def\@abssec#1{\vspace{.05in}\footnotesize \parindent .2in
{\bf #1. }\ignorespaces}

\graphicspath{{/EPSF/}{Figures/}}

\newtheorem{theorem}{Theorem}
\newtheorem{lemma}{Lemma}[section]

\def \Rm {\mathbb R}
\def \Nm {\mathbb N}

\newcommand{\eps}{\varepsilon}

\newcommand{\E}{\mathbb E}
\renewcommand{\P}{\mathbb P}
\newcommand{\dsum}{\displaystyle\sum}
\newcommand{\dint}{\displaystyle\int}
\newcommand{\dprod}{\displaystyle\prod}
\newcommand{\pdr}[2]{\dfrac{\partial{#1}}{\partial{#2}}}

\newcommand{\dr}[2]{\dfrac{d{#1}}{d{#2}}}

\newcommand{\bs}{\mathbf s}

\newcommand{\ban}{{\bar n}}

\newcommand{\btau}{\boldsymbol\tau}
\newcommand{\bxi}{\boldsymbol \xi}

\newcommand{\bzeta}{\boldsymbol \zeta}

\newcommand{\mg}{\mathfrak g}
\newcommand{\mG}{\mathfrak G}

\newcommand{\cout}[1]{}

\newcommand{\m}{{\mathfrak m}}
\newcommand{\n}{{\mathfrak n}}
\newcommand{\uu}{{\mathfrak u}}
\newcommand{\rr}{{\mathfrak r}}
\newcommand{\UU}{{\mathfrak U}}

\newcommand{\Ues}{{U_{\eps,s}}}

 \renewcommand{\arraystretch}{1.5}

\title{Homogenization with large spatial random potential}

\author{Guillaume Bal \thanks{Department of Applied Physics and 
        Applied Mathematics, Columbia University, 
        New York NY, 10027; gb2030@columbia.edu}}

\begin{document}
 
\maketitle


\begin{abstract}
  We consider the homogenization of parabolic equations with large
  spatially-dependent potentials modeled as Gaussian random fields. We
  derive the homogenized equations in the limit of vanishing
  correlation length of the random potential.  We characterize the
  leading effect in the random fluctuations and show that their
  spatial moments converge in law to Gaussian random variables.  Both
  results hold for sufficiently small times and in sufficiently large
  spatial dimensions $d\geq\m$, where $\m$ is the order of the spatial
  pseudo-differential operator in the parabolic equation. In dimension
  $d<\m$, the solution to the parabolic equation is shown to converge
  to the (non-deterministic) solution of a stochastic equation in the
  companion paper \cite{B-CMP-2-08}.  The results are then extended to
  cover the case of long range random potentials, which generate
  larger, but still asymptotically Gaussian, random fluctuations.
\end{abstract}
 

\renewcommand{\thefootnote}{\fnsymbol{footnote}}
\renewcommand{\thefootnote}{\arabic{footnote}}

\renewcommand{\arraystretch}{1.1}

\paragraph{keywords:}
Homogenization theory, partial differential equations with random
coefficients, Gaussian fluctuations, large potential, long range
correlations

\paragraph{AMS:} 35R60, 60H05, 35K15.



\section{Introduction}
\label{sec:intro}

Let $\m>0$ and $P(D)$ the pseudo-differential operator with symbol
$\hat p(\xi)=|\xi|^\m$. We consider the following evolution equation
in dimension $d\geq\m$:
\begin{equation}
  \label{eq:parabeps}
   \begin{array}{rcll}
  \Big(\pdr{}t + P(D) - \dfrac{1}{\eps^\alpha} q\big(\dfrac x\eps\big)
  \Big) u_\eps(t,x) &=&0,\qquad  & x\in\Rm^d,\quad t>0, \\
  u_\eps(0,x) &=& u_0(x), \quad& x\in\Rm^d .
   \end{array}
\end{equation}
Here, $u_0\in L^2(\Rm^d)$ and $q(x)$ is a mean zero stationary
Gaussian process defined on a probability space $(\Omega,\mathcal
F,\P)$. We assume that $q(x)$ has bounded and integrable correlation
function $R(x)=\E\{q(y)q(x+y)\}$, where $\E$ is the mathematical
expectation associated with $\P$, and bounded, continuous in the
vicinity of $0$, and integrable power spectrum $(2\pi)^d\hat
R(\xi)=\int_{\Rm^d} e^{-i\xi\cdot x} R(x)dx$ in the sense that
$\int_{\Rm^d\backslash B(0,1)} \hat R(\xi) |\xi|^{-\m}d\xi<\infty$.
The size of the potential is constructed so that the limiting solution
as $\eps\to0$ is different from the unperturbed solution obtained by
setting $q=0$.  The appropriate size of the potential is given by
\begin{equation}
  \label{eq:epsalpha}
   \eps^{\alpha} = \left\{ 
    \begin{matrix}
      \eps^{\frac{\m}2}|\ln\eps|^{\frac12} & d=\m, \\
      \eps^{\frac{\m}2} & d>\m.
    \end{matrix}\right.
\end{equation}

The potential is bounded $\P$-a.s. on bounded domains but is unbounded
$\P$-a.s. on $\Rm^d$. By using a method based on the Duhamel
expansion, we nonetheless obtain that for a sufficiently small time
$T>0$, the above equation admits a weak solution $u_\eps(t,\cdot)\in
L^2(\Omega\times\Rm^d)$ uniformly in time $t\in (0,T)$ and
$0<\eps<\eps_0$.

Moreover, as $\eps\to0$, the solution $u_\eps(t)$ converges strongly in
$L^2(\Omega\times\Rm^d)$ uniformly in $t\in (0,T)$ to its limit $u(t)$
solution of the following homogenized evolution equation
\begin{equation}
  \label{eq:parab}
   \begin{array}{rcll}
  \Big(\pdr{}t + P(D) -\rho
  \Big) u(t,x) &=&0,\qquad  & x\in\Rm^d,\quad t>0, \\
  u(0,x) &=& u_0(x), \quad & x\in\Rm^d ,
   \end{array}
\end{equation}
where the effective (non-negative) potential is given by
\begin{equation}
  \label{eq:rho}
  \rho =  \left\{ 
    \begin{array}{ll}
      \,\,c_d \hat R(0) & d=\m, \\[0mm]
      \dint_{\Rm^d} \dfrac{\hat R(\xi)}{|\xi|^\m} d\xi \quad & d>\m.
    \end{array}\right.
\end{equation}
Here, $c_d$ is the volume of the unit sphere $S^{d-1}$.  We denote
by $\mathcal G^\rho_t$ the propagator for the above equation,
which to $u_0(x)$ associates $\mathcal G^\rho_t u_0 (x) = u(t,x)$
solution of \eqref{eq:parab}.

We assume that the non-negative (by Bochner's theorem) power spectrum
$\hat R(\xi)$ is bounded by $f(|\xi|)$, where $f(r)$ is a positive,
bounded, radially symmetric, and integrable function in the sense that
$\int_1^\infty r^{d-1-\m}f(r)dr <\infty$. Then we have the
following result.
\begin{theorem}
  \label{thm:convergence} There exists a time $T=T(f)>0$ such that
  for all $t\in (0,T)$, there exists a solution $u_\eps(t)\in
  L^2(\Omega\times \Rm^d)$ uniformly in $0<\eps<\eps_0$.  Moreover,
  let us assume that $\hat R(\xi)$ is of class $\mathcal
  C^\gamma(\Rm^d)$ for some $0\leq \gamma\leq 2$ and let $u(t,x)$ be
  the unique solution in $L^2(\Rm^d)$ to \eqref{eq:parab}. Then, we
  have the convergence results
  \begin{equation}
  \label{eq:L2error}
  \begin{array}{rcl}
  \|(u_\eps-\uu_\eps)(t)\|_{L^2(\Omega\times\Rm^d)}
    &\lesssim& \eps^{\frac\beta2} 
  \|u_0\|_{L^2(\Rm^d)},  \\[2mm]
  \|(\uu_\eps-u)(t)\|_{L^2(\Rm^d)} &\lesssim& \eps^{\gamma\wedge\beta} 
  \|u_0\|_{L^2(\Rm^d)},
  \end{array}
  \end{equation}
  where $a\lesssim b$ means $a\leq Cb$ for some
  $C>0$, $a\wedge b=\min(a,b)$, where $\uu_\eps(t,\cdot)$ is a
  deterministic function in $L^2(\Rm^d)$ uniformly in time, and where
  we have defined
  \begin{equation}
  \label{eq:beta}
  \eps^\beta = \left\{ 
    \begin{array}{lc}
              |\ln\eps|^{-1} & d=\m, \\
              \eps^{d-\m} & \m<d<2\m, \\
              \eps^{\m}|\ln\eps| \quad& d=2\m, \\
              \eps^{\m} & d> 2\m.
    \end{array} \right.
  \end{equation}
  The Fourier transform $\UU_\eps(t,\xi)$ of the deterministic
  function $\uu_\eps(t,x)$ is determined explicitly in
  \eqref{eq:tildeUeps} below.
\end{theorem}
Note that the effective potential $-\rho$ is non-positive. The theorem
is valid for times $T$ such that $4T\rho_f<1$, where $\rho_f$ is
defined in lemma \ref{lem:bound} below by replacing $\hat R(\xi)$ by
$f(|\xi|)$ in the definition of $\rho$ in \eqref{eq:rho}.

The error term $u_\eps-u$ is dominated by deterministic components
when $\eps^{\gamma\wedge \beta}\gg\eps^{\frac{d-2\alpha}2}$ and by
random fluctuations when
$\eps^{\gamma\wedge\beta}\ll\eps^{\frac{d-2\alpha}2}$. In both
situations, the random fluctuations may be estimated as follows.  We
show that
\begin{equation}
  \label{eq:u1eps}
  u_{1,\eps}(t,x) = \dfrac{1}{\eps^{\frac{d-2\alpha}2}}
    \big(u_\eps-\E\{u_\eps\}\big)(t,x),
\end{equation}
converges weakly in space and in distribution to a Gaussian random
variable. More precisely, we have
\begin{theorem}
  \label{thm:fluct} 
  Let $M$ be a test function such that its Fourier transform $\hat
  M\in L^1(\Rm^d)\cap L^2(\Rm^d)$. Then we find that for all $t\in
  (0,T)$
  \begin{equation}
  \label{eq:conv}
  (u_{1,\eps}(t,\cdot),M) \xrightarrow{\,\eps\to0\,} 
   \dint_{\Rm^d} {\mathcal M}_t(x)  \sigma dW_x, \qquad 
   \mathcal M_t(x) = 
  \dint_0^t \mathcal G^\rho_s M(x) \mathcal G^\rho_{t-s} u_0(x) ds,
  \end{equation}
  where convergence holds in the sense of distributions, $dW_x$ is the
  standard multiparameter Wiener measure on $\Rm^d$ and $\sigma$ is
  the standard deviation defined by
  \begin{equation}
  \label{eq:sigmavar}
  \sigma^2 := (2\pi)^d\hat R(0) = \dint_{\Rm^d} \E\{q(0)q(x)\} dx.
  \end{equation}
\end{theorem}
This shows that the fluctuations of the solution are asymptotically
given by a Gaussian random variable, which is consistent with the
central limit theorem.


We observe a sharp transition in the behavior of $u_\eps$ at $d=\m$.
For $d<\m$, the following holds. The size of the potential that
generates an order $O(1)$ perturbation is now given by (see the last
inequality in lemma \ref{lem:bound})
\begin{displaymath}
  \eps^\alpha = \eps^{\frac d2}.
\end{displaymath}
Using the same methods as for the case $d\geq\m$, we may obtain that
$u_\eps(t)$ is uniformly bounded and thus converges weakly in
$L^2(\Omega\times\Rm^d)$ for sufficiently small times to a function
$u(t)$. The problem is addressed in \cite{B-CMP-2-08}, where it is
shown that $u(t)$ is the solution to the stochastic partial
differential equation in Stratonovich form
\begin{equation}
  \label{eq:stoch}
  \pdr{u}t + P(D) u + u\circ \sigma \dr Wx=0,
\end{equation}
with $u(0,x)=u_0(x)$ and $\frac{dW}{dx}$ d-parameter spatial white
noise ``density''. The above equation admits a unique solution that
belongs to $L^2(\Omega\times\Rm^d)$ locally uniformly in time.
Stochastic equations have also been analyzed in the case where
$d\geq\m$ (i.e., $d\geq2$ when $P(D)=-\Delta$), see
\cite{Hu-PA-02,NR-JFA-97}. However, our results show that such
solutions cannot be obtained as a limit in $L^2(\Omega\times\Rm^d)$ of
solutions corresponding to vanishing correlation length so that their
physical justification is more delicate.  In the case $d=1$ and $\m=2$
with $q(x)$ a bounded potential, we refer the reader to
\cite{PP-GAK-06} for more details on the above stochastic equation.

The above theorems \ref{thm:convergence} and \ref{thm:fluct} assume
short range correlations for the random potential. Mathematically,
this is modeled by an integrable correlation function, or equivalently
a bounded value for $\hat R(0)$. Longer range correlations may be
modeled by unbounded power spectra in the vicinity of the origin, for
instance by assuming that $\hat R(\xi)=h(\xi)\hat S(\xi)$, where $\hat
S(\xi)$ is bounded in the vicinity of the origin and $h(\xi)$ is a
homogeneous function of degree $-\n$ for some $\n>0$.  Provided that
$d>\m+\n$ so that $\rho$ defined in \eqref{eq:rho} is still bounded,
the results of theorems \ref{thm:convergence} and \ref{thm:fluct} may
be extended to the case of long range fluctuations. We refer the
reader to theorem \ref{thm:fracBM} in section \ref{sec:longrange}
below for the details. The salient features of the latter result is
that the convergence properties stated in theorem
\ref{thm:convergence} still hold with $\beta$ replaced by $\beta-\n$
and that the random fluctuations are now asymptotically Gaussian
processes of amplitude of order $\eps^{\frac{d-\m-\n}2}$. Moreover,
they may conveniently be written as stochastic integrals with respect
to some multiparameter fractional Brownian motion in place of the
Wiener measure appearing in \eqref{eq:conv}.

Let us also mention that all the result stated here extend to the
Schr\"odinger equation, where $\frac{\partial}{\partial t}$ is
replaced by $i\frac{\partial}{\partial t}$ in \eqref{eq:parabeps}.  We
then verify that $-\rho$ in \eqref{eq:parab} is replaced by $\rho$
so that the homogenized equation is given by
\begin{displaymath}
  \Big(i\pdr{}t + P(D) +\rho
  \Big) u(t,x) \,=\,0.
\end{displaymath}
The main effect of the randomness is therefore a phase shift of the
quantum waves as they propagate through the random medium. Because the
semigroup associated to the free evolution of quantum waves does not
damp high frequencies as efficiently as for the parabolic equation
\eqref{eq:parabeps}, some additional regularity assumptions on the
initial condition are necessary to obtain the limiting behaviors
described in theorems \ref{thm:convergence} and \ref{thm:fluct}. We do
not consider the case of the Schr\"odinger equation further here.

The rest of the paper is structured as follows. Section
\ref{sec:existence} recasts \eqref{eq:parabeps} as an infinite Duhamel
series of integrals in the Fourier domain. The cross-correlations of
the terms appearing in the series are analyzed by calculating
moments of Gaussian variables and estimating the contributions of
graphs similar to those introduced in \cite{Erdos-Yau2,LS-ARMA-07}.
These estimates allow us to construct a solution to
\eqref{eq:parabeps} in $L^2(\Omega\times\Rm^d)$ uniformly in time for
sufficiently small times $t\in (0,T)$. The maximal time $T$ of
validity of the theory depends on the power spectrum $\hat R(\xi)$.
The estimates on the graphs are then used in section \ref{sec:limits}
to characterize the limit and the leading random fluctuations of the
solution $u_\eps(t,x)$. The extension of the results to long range
correlations is presented in section \ref{sec:longrange}.

The analysis of \eqref{eq:parabeps} and of similar operators has been
performed for smaller potentials than those given in
\eqref{eq:epsalpha} in e.g. \cite{B-CLH-08,FOP-SIAP-82} when $u_\eps$
converges strongly to the solution of the unperturbed equation (with
$q\equiv0$). The results presented in this paper may thus be seen as
generalizations to the case of sufficiently strong potentials so that
the unperturbed solution is no longer a good approximation of
$u_\eps$. The analysis presented below is based on simple estimates
for the Feynman diagrams corresponding to Gaussian random potentials
and does not extend to other potentials such as Poisson point
potentials, let alone potentials satisfying some mild mixing
conditions.  Extension to other potentials would require more
sophisticated estimates of the graphs than those presented here or a
different functional setting than the $L^2(\Omega\times\Rm^d)$ setting
considered here. For related estimates on the graphs appearing in
Duhamel expansion, we refer the reader to e.g.
\cite{Chen-JSP-05,Erdos-Yau2,LS-ARMA-07}.

\section{Duhamel expansion and existence theory}
\label{sec:existence}

Since $q(x)$ is a stationary mean zero Gaussian random field, it
admits the following spectral representation
\begin{equation}
  \label{eq:spectq}
  q(x) = \dfrac{1}{(2\pi)^d} \dint_{\Rm^d} e^{i\xi\cdot x}
   \hat Q(d\xi),
\end{equation}
where $\hat Q(d\xi)$ is the complex spectral process such that
\begin{displaymath}
  \E\Big\{ \dint_{\Rm^d} f(\xi) \hat Q(d\xi)
    \overline{\dint_{\Rm^d} g(\xi) \hat Q(d\xi)} \Big\}
  = \dint_{\Rm^d} f(\xi) \bar g(\xi) (2\pi)^d\hat R(\xi) d\xi,
\end{displaymath}
for all $f$ and $g$ in $L^2(\Rm^d;\hat R(\xi)d\xi)$ with the power
spectrum and correlation function of $q$ respectively defined by
\begin{equation}
  \label{eq:RhatR}
  0\leq (2\pi)^d \hat R(\xi) = \dint_{\Rm^d} e^{-i\xi\cdot x} R(x) dx,\qquad
  R(x) = \E\{q(y)q(x+y)\}. 
\end{equation}
In the sequel, we write $\hat Q(d\xi)\equiv\hat q(\xi) d\xi$ so that
$\E\{\hat q(\xi)\hat q(\zeta)\}=\hat R(\xi)\delta(\xi+\zeta)$ and
$\E\{\hat q(\xi)\overline{\hat q(\zeta)}\}=\hat
R(\xi)\delta(\xi-\zeta)$.

\subsection{Duhamel expansion}
\label{sec:duhamel}

Let us introduce $\hat q_\eps(\xi)=\eps^{d-\alpha}\hat q(\eps\xi)$,
the Fourier transform of $\eps^{-\alpha}q(\frac{x}\eps)$.  We may now
recast the parabolic equation \eqref{eq:parabeps} as
\begin{equation}
  \label{eq:parabFD}
  \big(\pdr{}t + \xi^\m \big) \hat u_\eps = \hat q_\eps * \hat u_\eps,
\end{equation}
with $\hat u_\eps(0,\xi)=\hat u_0(\xi)$, where
\begin{displaymath}
  \hat q_\eps * \hat u_\eps (t,\xi) = 
  \dint_{\Rm^d} \hat u_\eps(t,\xi-\zeta) \hat Q_\eps(d\zeta)
   \equiv \dint_{\Rm^d} \hat u_\eps(t,\xi-\zeta) \hat q_\eps(\zeta)d\zeta.
\end{displaymath}
Here and below, we use the notation $\xi^\m=|\xi|^\m$. After integration in
time, the above equation becomes
\begin{equation}
  \label{eq:intFD}
  \hat u_\eps(t,\xi) = e^{-t\xi^\m} \hat u_0(\xi) + \dint_0^t
   e^{-s \xi^\m} \dint_{\Rm^d} \hat q_\eps(\xi-\xi_1) \hat u_\eps(t-s,\xi_1)
   d\xi_1 ds.
\end{equation}
This allows us to write the formal Duhamel expansion
\begin{eqnarray}
  \label{eq:exp}
  \hat u_\eps(t,\xi) &=& \dsum_{n\in\Nm} \hat u_{n,\eps}(t,\xi),\\
  \label{eq:hatuneps}
  \hat u_{n,\eps}(t,\xi_0) &=& \dint_{\Rm^{nd}} \dprod_{k=0}^{n-1}
   \dint_0^{t_k(\bs)} e^{-\xi_k^\m s_k}  e^{-(t-\sum_{k=0}^{n-1} s_k)\xi_n^\m}
    \dprod_{k=0}^{n-1} \hat q_\eps(\xi_k-\xi_{k+1})
   \hat u_0(\xi_n)  d\bs d\bxi.
\end{eqnarray}
Here, we have introduced the following notation:
\begin{displaymath}
  \bs=(s_0,\ldots,s_{n-1}),\,\, t_k(\bs) = t-s_0-\ldots - s_{k-1},\,\,
  t_0(\bs)=t,\,\, d\bs=\prod_{k=0}^{n-1}ds_k,\,\, d\bxi=\prod_{k=1}^n d\xi_k.
\end{displaymath}

We now show that for sufficiently small times, the expansion
\eqref{eq:exp} converges (uniformly for all $\eps$ sufficiently small)
in the $L^2(\Omega\times \Rm^d)$ sense. Moreover, the $L^2$ norm of
$u_\eps(t)$ is bounded by the $L^2(\Rm^d)$ norm of $\hat u_0$, which
gives us an a priori estimate for the solution.  The convergence
results are based on the analysis of the following moments
\begin{equation}
  \label{eq:Ueps}
  U^{n,m}_\eps(t,\xi,\zeta) = \E\{\hat u_{\eps,n}(t,\xi) 
    \overline{\hat u_{\eps,m}}(t,\zeta)\},
\end{equation}
which, thanks to \eqref{eq:hatuneps}, are given by
\begin{displaymath}
  \begin{array}{l}
      \dint_{\Rm^{d(n+m)}} \prod_{k=0}^{n-1}
  \dint_0^{t_k(\bs)}  \prod_{l=0}^{m-1} \dint_0^{t_l(\btau)} 
  e^{- s_k\xi_k^\m}
  e^{-(t-\sum_{k=0}^{n-1} s_k)\xi_n^\m}
  e^{- \tau_l\zeta_l^\m}
  e^{-(t-\sum_{l=0}^{m-1} \tau_l)\zeta_m^\m} \\
  \E\Big\{\dprod_{k=0}^{n-1} \dprod_{l=0}^{m-1}
   \hat q_\eps(\xi_k-\xi_{k+1})\bar {\hat q}_\eps(\zeta_l-\zeta_{l+1})\Big\}
    \hat u_0(\xi_n)\bar {\hat u}_0(\zeta_m) \, d\bs d\btau d\bxi d\bzeta.
  \end{array}
\end{displaymath}
Let us introduce the notation $s_{n}(\bs)=t_n(\bs)=t-\sum_{k=0}^{n-1}
s_k$ and $\tau_{m}(\btau)=t_m(\btau)=t-\sum_{l=0}^{m-1}\tau_l$.  We
also define $\xi_{n+k+1}=\zeta_{m-k}$ and $s_{n+k+1}=\tau_{m-k}$ for
$0\leq k\leq m$. Since $q_\eps$ is real-valued, we find that
\begin{displaymath}
   U_\eps^{n,m}(t,\xi_0,\xi_{n+m+1}) =
  \dint \prod_{k=0}^{n+m+1} e^{-s_k\xi_k^\m}
   \E\Big\{\prod_{k=0,k\not=n}^{n+m} \hat q_\eps(\xi_k-\xi_{k+1}) \Big\}
   \hat u_0(\xi_n) \bar {\hat u}_0(\xi_{n+1}) d\bs d\bxi,
\end{displaymath}
where the domain of integration in the $s$ and $\xi$ variables is
inherited from the previous expression.  Note that no integration is
performed in the variables $s_n(\bs)$ and $s_{n+1}(\btau)$. The
integral may be recast as
\begin{displaymath}
  \dint \prod_{k=0}^{n+m+1} e^{-s_k\xi_k^2}
   \E\Big\{\prod_{k=0,k\not=n}^{n+m} \hat q_\eps(\xi_k-\xi_{k+1}) \Big\}
   \hat u_0(\xi_n) \bar {\hat u}_0(\xi_{n+1}) 
   \delta(t-\sum_{k=0}^{n}s_k)
   \delta(t-\sum_{k=n+1}^{n+m+1}s_k) d\bs d\bxi,
\end{displaymath}
where the integrals in all the $s_k$ variables for $0\leq k\leq n+m+1$
are performed over $(0,\infty)$.  The $\delta$ functions ensure that
the integration is equivalent to the one presented above. The latter
form is used in the proof of lemma \ref{lem:timeint} below.

We need to introduce additional notation. The moments of $\hat u_{\eps,n}$
are defined as
\begin{equation}
  \label{eq:Uepsn}
  U_\eps^n(t,\xi) = \E\{\hat u_{\eps,n}(t,\xi) \}.
\end{equation}
We also introduce the following covariance function
\begin{equation}
  \label{eq:Vepsnm}
   V^{n,m}_\eps(t,\xi,\zeta) = {\rm cov}(\hat u_{\eps,n}(t,\xi), 
   \hat u_{\eps,m}(t,\zeta)) =
   U^{n,m}_\eps(t,\xi,\zeta) - U_\eps^n(t,\xi)
   \overline{U_\eps^m(t,\zeta)}.
\end{equation}
These terms allow us to analyze the convergence properties of the
solution $\hat u_\eps(t,\xi)$. Let $\hat M(\xi)$ be a smooth
(integrable and square integrable is sufficient) test function on
$\Rm^d$. We introduce the two random variables
\begin{eqnarray}
  \label{eq:energy}
  I_\eps(t) &=& \dint_{\Rm^d} |\hat u_\eps(t,\xi)|^2 d\xi\\
  \label{eq:Xeps}
  X_\eps(t) &=& \dint_{\Rm^d} \hat u_\eps(t,\xi) \overline{\hat M}(\xi)d\xi.
\end{eqnarray}

\subsection{Summation over graphs}
\label{sec:graphs}

We now need to estimate moments of the Gaussian process $\hat q_\eps$.
The expectation in $U_\eps^{n,m}$ vanishes unless there is
$\ban\in\Nm$ such that $n+m=2\ban$ is even. The expectation of a
product of Gaussian variables has an explicit structure written as a
sum over all possible products of pairs of indices of the form
$\xi_k-\xi_{k+1}$.  The moments are thus given as a sum of products of
the expectation of {\em pairs} of terms $\hat
q_\eps(\xi_k-\xi_{k+1})$, where the sum runs over all possible
pairings. We define the {\bf pair} $(\xi_k,\xi_l)$, $1\leq k<l$, as
the contribution in the product given by
\begin{displaymath}
  \E\{\hat q_\eps(\xi_{k-1}-\xi_{k})\hat q_\eps(\xi_{l-1}-\xi_{l})\}
   = \eps^{d-2\alpha} \hat R(\eps(\xi_k-\xi_{k-1}))
  \delta(\xi_k-\xi_{k-1}+\xi_l-\xi_{l-1}).
\end{displaymath}
We have used here the fact that $\hat R(-\xi)=\hat R(\xi)$.

The number of pairings in a product of $n+m=2\ban$ terms (i.e., the
number of allocations of the set $\{1,\ldots,2\ban\}$ into $\ban$
unordered pairs) is equal to
\begin{displaymath}
  \dfrac{(2\ban-1)!}{2^{\ban-1}(\ban-1)!} = \dfrac{(2\ban)!}{\ban!2^\ban}
  =(2n-1)!!.
\end{displaymath}
There is consequently a very large number of terms appearing in
$U_\eps^{n,m}(t,\xi_0,\xi_{n+m+1})$. In each instance of the pairings,
we have $\ban$ terms $k$ and $\ban$ terms $l\equiv l(k)$.  Note that
$l(k)\geq k+1$. We denote by {\bf simple pairs} the pairs such that
$l(k)=k+1$, which thus involve a delta function of the form
$\delta(\xi_{k+1}-\xi_{k-1})$.

\begin{figure}[htbp]
  \centering
  \includegraphics[height=3.5cm]{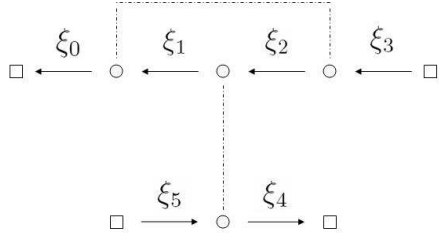}
  \caption{Graph with $n=3$ and $m=1$ corresponding to the pairs $(\xi_1,\xi_3)$ and $(\xi_2,\xi_5)$ and the delta functions $\delta(\xi_1-\xi_0+\xi_3-\xi_2)$
    and $\delta(\xi_2-\xi_1+\xi_5-\xi_4)$.}
  \label{fig:1}
\end{figure}
The collection of pairs $(\xi_k,\xi_{l(k)})$ for $\ban$ values of $k$
and $\ban$ values of $l(k)$ constitutes a graph $\mg\in \mG$
constructed as follows; see Fig.\ref{fig:1} and \cite{Erdos-Yau2}.
The upper part of the graph with $n$ bullets represents $\hat
u_{\eps,n}$ while the lower part with $m$ bullets represents
$\overline{\hat u_{\eps,m}}$. The two squares on the left of the graph
represent the variables $\xi_0$ and $\xi_{n+m+1}$ in
$U_\eps^{n,m}(t,\xi_0,\xi_{n+m+1})$ while the squares on the right
represent $\hat u_0(\xi_n)$ and $\bar {\hat u}_0(\xi_{n+1})$. The
dotted pairing lines represent the pairs of the graph $\mg$.  Here,
$\mG$ denotes the collection of all possible
$|\mG|=\frac{(2\ban-1)!}{2^{\ban-1}(\ban-1)!}$ graphs that can be
constructed for a given $\ban$.

We denote by $A_0=A_0(\mg)$ the collection of the $\ban$ values of $k$
and by $B_0=B_0(\mg)$ the collection of the $\ban$ values of $l(k)$.
We then find that
\begin{displaymath}
  \E\Big\{\prod_{k=0,k\not=n}^{n+m} \hat q_\eps(\xi_k-\xi_{k-1}) \Big\}
 = \dsum_{\mg\in\mG} \dprod_{k\in A_0(\mg)}
   \eps^{d-2\alpha} \hat R(\eps(\xi_k-\xi_{k-1}))
  \delta(\xi_{k}-\xi_{k-1}+\xi_{l(k)}-\xi_{l(k)-1}).
\end{displaymath}
This provides us with an explicit expression for
$U_\eps^{n,m}(t,\xi_0,\xi_{n+m+1})$ as a summation over all possible
graphs generated by moments of Gaussian random variables. We need to
introduce several classes of graphs. 

We say that the graph has a {\bf crossing} if there is a $k\leq n$
such that $l(k)\geq n+2$. We denote by $\mG_c\subset \mG$ the set of
graphs with at least one crossing and by $\mG_{nc}=\mG\backslash\mG_c$
the {\bf non-crossing} graphs. We observe that
$V^{n,m}_\eps(t,\xi_0,\xi_{n+m+1})$ is the sum over the crossing
graphs and that $U_\eps^n(t,\xi_0) \overline{U_\eps^m(t,\xi_{n+m+1})}$
is the sum over the non-crossing graphs in
$U^{n,m}_\eps(t,\xi_0,\xi_{n+m+1})$.

The unique graph $\mg_s$ with only simple pairs is called the {\bf
  simple graph} and we define $\mG_{ns}=\mG\backslash\mg_s$.  We
denote by $\mG_{cs}$ the {\bf crossing simple graphs} with only simple
pairs except for exactly one crossing. The complement of $\mG_{cs}$ in
the crossing graphs is denoted by $\mG_{cns}=\mG_c\backslash\mG_{cs}$.

As we shall see, only the simple graph $\mg_s$ contributes an $O(1)$
term in the limit $\eps\to0$ and only the graphs in $\mG_{cs}$
contribute to the leading order $O(\eps^{\frac12(d-2\alpha)})$ in the
fluctuations of $\hat u_\eps$.

The graphs are defined similarly in the calculation of
$U_\eps^n(t,\xi_0)$ in \eqref{eq:Uepsn} for $n=2\ban$ and $m=0$,
except that crossing graphs have no meaning in such a context. A
summation over $k\in A_0(\mg)$ of all the arguments
$\xi_{k}-\xi_{k-1}+\xi_{l(k)}-\xi_{l(k)-1}$ of the $\delta$ functions
shows that the last delta function may be replaced without modifying
the integral in $U_\eps^n(t,\xi_0)$ by $\delta(\xi_0-\xi_n)$.

This allows us to summarize the above calculations as follows:
\begin{equation}
  \label{eq:Uepsmn}
  \begin{array}{l}
 U_\eps^{n,m}(t,\xi_0,\xi_{n+m+1}) =
  \dint \prod_{k=0}^{n+m+1} e^{-s_k\xi_k^\m}
      \hat u_0(\xi_n) \bar {\hat u}_0(\xi_{n+1}) \dsum_{\mg\in\mG} \\
 \qquad  \dprod_{k\in A_0(\mg)}
   \eps^{d-2\alpha} \hat R(\eps(\xi_k-\xi_{k-1}))
  \delta(\xi_{k}-\xi_{k-1}+\xi_{l(k)}-\xi_{l(k)-1})
   d\bs d\bxi.
  \end{array}
\end{equation}
Similarly,
\begin{equation}
  \label{eq:Uepsm}
  \begin{array}{l}
 U_\eps^{n}(t,\xi_0) 
  = \hat u_0(\xi_0)
    \dint \prod_{k=0}^{n} e^{-s_k\xi_k^\m}\dsum_{\mg\in\mG}  \\\qquad
  \dprod_{k\in A_0(\mg)}
   \eps^{d-2\alpha} \hat R(\eps(\xi_k-\xi_{k-1}))
  \delta(\xi_{k}-\xi_{k-1}+\xi_{l(k)}-\xi_{l(k)-1})
   d\bs d\bxi. \!\!\!\!\!\!\!\!\!\!\!\! 
  \end{array}
\end{equation}

\subsection{Analysis of crossing graphs}

We now analyze the influence of the crossing graphs on $I_\eps(t)$ and
$X_\eps(t)$ defined in \eqref{eq:energy} and \eqref{eq:Xeps},
respectively, for sufficiently small times. We obtain from
\eqref{eq:Vepsnm} and \eqref{eq:Uepsmn} that
\begin{equation} \label{eq:Vepsmnint}
  \begin{array}{l}
  V^{n,m}_\eps(t,\xi_0,\xi_{n+m+1})
   =\dsum_{\mg\in\mG_c} 
  \dint \prod_{k=0}^{n+m+1} e^{-s_k\xi_k^\m}
      \hat u_0(\xi_n) \bar {\hat u}_0(\xi_{n+1}) \\
 \qquad\quad   \dprod_{k\in A_0(\mg)}
   \eps^{d-2\alpha} \hat R(\eps(\xi_k-\xi_{k-1}))
  \delta(\xi_{k}-\xi_{k-1}+\xi_{l(k)}-\xi_{l(k)-1})
   \,d\bs \,d\bxi,
  \end{array}
\end{equation}
involves the summation over the crossing graphs $\mG_c$. Let us
consider a graph $\mg\in\mG_c$ with $M$ crossing pairs, $M\geq1$.
Crossing pairs are defined by $k\leq n$ and $l(k)\geq n+2$.  Denote by
$(\xi_{q_m},\xi_{l(q_m)})$, $1\leq m\leq M$ the crossing pairs and
define $Q=\max_m\{q_m\}$. By summing the arguments inside the delta
functions for all $k\leq n$, we observe that the last of these
delta functions may be replaced by 
\begin{displaymath}
  \delta(\xi_0-\xi_n + \dsum_{m=1}^M \xi_{q_m}-\xi_{q_m-1}).
\end{displaymath}
Similarly, by summing over all pairs with $k\geq n+2$, we obtain that
the last of these delta functions may be replaced by
\begin{displaymath}
  \delta(\xi_{n+1}-\xi_{n+m+1}+ \dsum_{m=1}^M 
  \xi_{l(q_m)}-\xi_{l(q_m)-1}).
\end{displaymath}
The product of the latter two delta functions is then equivalent to
\begin{displaymath}
  \delta(\xi_{n+m+1}-\xi_{n+1}+\xi_n-\xi_0)
   \delta(\xi_Q-\xi_{Q-1} + \xi_0-\xi_n + \dsum_{m=1}^{M-1}
   \xi_{q_m}-\xi_{q_m-1}).
\end{displaymath}
The analysis of the contributions of the crossing graphs is slightly
different for the energy in \eqref{eq:energy} and for the spatial
moments in \eqref{eq:Xeps}. We start with the energy.

\paragraph{Analysis of the crossing terms in $I_\eps(t)$.}  
We evaluate the expression for $|V^{n,m}_\eps(t,\xi_0,\xi_0)|$ in
\eqref{eq:Vepsmnint} at $\xi_{n+m+1}=\xi_0$ and integrate in the
$\xi_0$ variable over $\Rm^d$.  Let us define $A'=A_0\backslash\{Q\}$.
For each $k\in A'\cup\{0\}$, we perform the change of variables
$\xi_k\to\frac{\xi_k}\eps$.  We then define
\begin{equation}
  \label{eq:xieps}
  \xi_k^\eps = \left\{
    \begin{array}{ll}
          \xi_k & k\not\in A'\cup\{0\} \\
          \frac{\xi_k}\eps & k\in A'\cup\{0\}.
    \end{array}\right.
\end{equation}
Note that $\xi_n=\xi_{n+1}$ since $\xi_{n+m+1}=\xi_0$.  This allows us
to obtain that
\begin{equation}
  \label{eq:Vepsmn2}
  \begin{array}{l}
 \dint_{\Rm^d} |V_\eps^{n,m}(t,\xi_0,\xi_{0})|d\xi_0  \leq
  \dsum_{\mg\in\mG_c}
  \dint   e^{-(s_0+s_{n+m+1})\eps^{-\m}\xi_0^\m}
  \prod_{k=1}^{n+m} e^{-s_k(\xi_k^\eps)^\m}
      |\hat u_0(\xi_n)|^2  \\
 \qquad  \dprod_{k\in A'(\mg)}
   \eps^{-2\alpha} \hat R(\xi_k-\eps\xi^\eps_{k-1})
  \delta(\frac{\xi_{k}}\eps-\xi^\eps_{k-1}+\xi_{l(k)}-\xi^\eps_{l(k)-1})\\
  \qquad \eps^{-2\alpha} \hat R(\xi_0-\eps\xi_n + 
      \dsum_{m=1}^M\xi_{q_m}-\eps\xi^\eps_{q_m-1})
   \delta(\xi_{n+1}-\xi_n)
   d\bs d\bxi.
   \end{array}
\end{equation}
Here $d\bxi$ also includes the integration in the variable $\xi_0$.
The estimates for $V^{n,m}_\eps$ here and in subsequent sections rely
on integrating selected time variables. All estimates are performed as
the following lemma indicates.
\begin{lemma}
  \label{lem:timeint}
  Let $t>0$ given and consider an integral of the form
  \begin{equation}
    \label{eq:intnm1}
    I_{n-1} = \dprod_{k=0}^{n-1}\dint_0^{t_k(\bs)}
     \Big(\prod_{k=0}^{n-1} f_k(s_k)\Big)  
    \prod_{k=0}^{n-1} ds_k,
  \end{equation}
  where $0\leq f_k(\bs)\leq 1$ for $0\leq k\leq n$ and assume that
  $\int_0^t f_{n-1}(s_{n-1})ds_{n-1} \leq h\wedge t$. Then
  \begin{equation}
    \label{eq:intnm2}
    I_{n-2} \leq (h\wedge t) I_{n-1}.
  \end{equation}
  Moreover, let ${\mathfrak s}$ be a permutation of the indices $0\leq k\leq
  n-1$.  Define $I^{\mathfrak s}_{n-1}$ as $I_{n-1}$ with $f_k$ replaced by
  $f_{{\mathfrak s}(k)}$. Then $I^{\mathfrak s}_{n-1}=I_{n-1}$.  
  
  Using the above result with the permutation leaving all indices
  fixed except ${\mathfrak s}(n-1)=K$ and ${\mathfrak s}(K)=n-1$ for
  some $0\leq K\leq n-2$ allows us to estimate $I_{n-1}$ by
  integrating in the $K$th variable.
\end{lemma}
\begin{proof}
  The derivation of \eqref{eq:intnm2} is immediate. We also calculate
  \begin{displaymath}
   \begin{array}{rclrcl}
    I_{n-1} &=&  \dint_{\Rm_+^{n+1}}
    \Big(\prod_{k=0}^{n-1} f_k(s_k)\Big) \delta\big(t-\dsum_{k=0}^n s_k\big)
    \prod_{k=0}^n ds_k\\
    &=&  \dint_{\Rm_+^{n+1}}
    \Big(\prod_{k=0}^{n-1} f_{{\mathfrak s}(k)}(s_{{\mathfrak s}(k)})\Big)
       \delta\big(t-\dsum_{k=0}^n s_{{\mathfrak s}(k)}\big)  
    \prod_{k=0}^n ds_{k} \\
    &=&  \dint_{\Rm_+^{n+1}}
    \Big(\prod_{k=0}^{n-1} f_{{\mathfrak s}(k)}(s_k)\Big)
    \delta\big(t-\dsum_{k=0}^n s_{k}\big)
    \prod_{k=0}^n ds_k &=& I^{\mathfrak s}_{n-1}.
   \end{array}
  \end{displaymath}
\end{proof}
Note that $e^{-s_{n}(\bs)(\xi_\eps^n)^\m}$ and
$e^{-s_{n+1}(\bs)(\xi_\eps^{n+1})^\m}$ are bounded by $1$. We now
estimate the integrals in the variables $s_0$, $s_{n+m+1}$, and $s_k$
for $k\in A'$ in \eqref{eq:Vepsmn2}. Note that $n+1$ cannot belong to
$A'$ and that $n$ does not belong to $A'$ either since either $n=Q$
(last crossing) or $n\in B_0$ is a receiving end of the pairing line
$k\to l(k)$. Each integral is bounded by:
\begin{equation}
   \label{eq:bdtime}
  \dint_0^{\tau\wedge t} e^{-s \eps^{-\m}\xi^\m} ds \leq
  \dfrac{\eps^\m}{\xi^\m} \wedge t.
\end{equation}
The remaining exponential terms $e^{-s_k(\xi^\eps_k)^\m}$ are bounded
by $1$. 
Using lemma \ref{lem:timeint}, this allows us to obtain that
\begin{displaymath}
  \begin{array}{l}
  \dint_{\Rm^d} |V_\eps^{n,m}(t,\xi_0,\xi_{0})| d\xi_0
   \leq  \dsum_{\mg\in\mG_c} \Big(\dint d\tilde\bs \Big) 
  \dint   |\hat u_0(\xi_n)|^2  \\ 
 \qquad \dprod_{k\in A'(\mg)}
    \eps^{-2\alpha} \Big(\dfrac{\eps^{\m}}{\xi_k^\m} \wedge t\Big)
  \hat R(\xi_k-\eps\xi^\eps_{k-1})
  \delta(\frac{\xi_{k}}\eps-\xi^\eps_{k-1}+\xi_{l(k)}-\xi^\eps_{l(k)-1})\\
  \qquad  \eps^{-2\alpha} \Big(\dfrac{\eps^{\m}}{\xi_0^\m} \wedge t\Big)^2
    \hat R(\xi_0-\eps\xi_n + 
      \dsum_{m=1}^M\xi_{q_m}-\eps\xi^\eps_{q_m-1})
   \delta(\xi_{n+1}-\xi_n) \,d\bxi.
   \end{array}
\end{displaymath}
Here, $d\tilde\bs$ corresponds to the integration in the remaining
time variables $s_k$ for $k\not\in A'\cup\{n+m+1\}$. There are
$2\ban-1-(\ban +1)=\ban-2$ such variables.  Note the square on the
last line, which comes from integrating in both variables $s_0$ and
$s_{n+m+1}$. 

The delta functions allow us to integrate in the variables
$\xi_{l(k)}$ for $k\in A'(\mg)$ and the initial condition $\hat
u_0(\xi_n)$ in the variable $\xi_n$.  Thanks to lemma
\ref{lem:bound} below, the power spectra allow us to integrate in
the remaining variables in $A'\cup\{0\}$. The integrals in the
variables in $A'$ are all bounded by $\rho_f$ defined in lemma
\ref{lem:bound} whereas the integral in $\xi_0$ results in a
bound equal to $\eps^\beta\rho_f$, where $\eps^\beta$ is defined in
\eqref{eq:beta}. As a consequence, we have the bound
\begin{displaymath}
  \dint_{\Rm^d} |V_\eps^{n,m}(t,\xi_0,\xi_{0})| d\xi_0
   \leq  \dsum_{\mg\in\mG_c}
  \Big(\dint d\tilde\bs \Big)  \rho_f^{\ban-1} \|\hat u_0\|^2
   \rho_f\eps^\beta =\dsum_{\mg\in\mG_c}
  \Big(\dint d\tilde\bs \Big) \rho_f^{\ban} \eps^\beta\|\hat u_0\|^2.
\end{displaymath}
Using Stirling's formula, we find that
$|\mG_c|<\frac{(2\ban-1)!}{2^{\ban-1}(\ban-1)!}$ is bounded by
$(\frac{2\ban}{e})^{\ban}$. It remains to evaluate
the integrals in time. We verify that
\begin{equation} \label{eq:sumtime}
  \prod_{k=0}^{n-1} \dint_0^{t_k(\bs)} ds_0\cdots ds_{n-1}
  =\dfrac{t^n}{n!},\qquad 
   t_k(\bs) = t-s_0-\ldots - s_{k-1}.
\end{equation}
 Let $\bar p=\bar p(\mg)$ be the number of $s_k$ for
$k\leq n$ in $\tilde\bs$ and $\bar q=\bar q(\mg)$ be the number of
$s_k$ for $k\geq n+1$ in $\tilde\bs$, with $\bar p+\bar q=\ban-1$.
Using \eqref{eq:sumtime}, we thus find that 
\begin{displaymath}
  \Big(\dint d\tilde\bs \Big) = \dfrac{t^{\bar p}}{\bar p!} 
   \dfrac{t^{\bar q}}{\bar q!} 
  = \dfrac{t^{\ban-1}}{( \ban-1)!} {\ban-1 \choose \bar p}
  \leq t^{\ban-1} \Big(\dfrac{\ban-1}{2e}\Big)^{-\ban+1}
   \leq t^{\ban-1} \ban \Big(\dfrac{\ban}{2e}\Big)^{-\ban}
\end{displaymath}
using Stirling's formula. This shows that
\begin{equation} \label{eq:boundgraphs}
  \dsum_{\mg\in\mG_c} \Big(\dint d\tilde\bs \Big) \leq
  \dfrac{\ban}{T} (4\rho_f T)^\ban,
\end{equation}
uniformly for $t\in (0,T)$. We thus need to choose $T$ sufficiently
small so that $4\rho_f T<1$. Then, for $\rr$ such that $4\rho_f T<\rr^2<1$, 
we find that 
\begin{equation} \label{eq:bdVeps1}
 \dint |V_\eps^{n,m}(t,\xi_0,\xi_{0})| d\xi_0
   \leq C \rr^{n+m} \eps^\beta \|\hat u_0\|^2,
\end{equation}
for some positive constant $C$. It remains to sum over $n$ and $m$ 
to obtain that 
\begin{equation}
  \label{eq:boundI}
  \big|\E\{I_\eps(t)\}-\dint_{\Rm^d} \E\{\hat u_\eps(t,\xi)\}^2 d\xi
  \big| \leq \dfrac{C}{\rr^2} \eps^\beta \|\hat u_0\|^2.
\end{equation}
We shall analyze the non-crossing terms generating $|\E\{\hat
u_\eps(t,\xi)\}|^2$ shortly. Before doing so, we analyze the influence
of the crossing terms on $X_\eps$. We can verify that the error term
$\eps^\beta$ in \eqref{eq:boundI} is optimal, for instance by looking
at the contribution of the graph with $n=m=1$.

\paragraph{Analysis of the crossing terms in $X_\eps$.}
It turns out that the contribution of the crossing terms is smaller
for the moment $X_\eps$ than it is for the energy $I_\eps$. More
precisely, we show that the smallest contribution to the variance of
$X_\eps$ is of order $\eps^{d-2\alpha}$ for graphs in $\mG_{cs}$ and
of order $\eps^{d-2\alpha+\beta}$ for the other crossing graphs.

We come back to \eqref{eq:Vepsmnint} and this time perform the change
of variables $\xi_k\to\frac{\xi_k}\eps$ for $k\in A'$ only.  We
re-define
\begin{equation}
  \label{eq:xieps2}
  \xi_k^\eps = \left\{
    \begin{array}{ll}
          \xi_k & k\not\in A' \\
          \frac{\xi_k}\eps & k\in A',
    \end{array}\right.
\end{equation}
and find that
\begin{equation}
  \label{eq:Vepsmn3}
  \begin{array}{l}
  V_\eps^{n,m}(t,\xi_0,\xi_{n+m+1}) =\dsum_{\mg\in\mG_c}
  \dint \prod_{k=0}^{n+m+1} e^{-s_k(\xi_k^\eps)^\m}
      \hat u_0(\xi_n) \bar {\hat u}_0(\xi_{n+1}) \\
 \qquad  \dprod_{k\in A'(\mg)}
   \eps^{-2\alpha} \hat R(\xi_k-\eps\xi^\eps_{k-1})
  \delta(\frac{\xi_{k}}\eps-\xi^\eps_{k-1}+\xi_{l(k)}-\xi^\eps_{l(k)-1})\\
  \qquad \eps^{d-2\alpha} \hat R(\eps(\xi_Q-\xi^\eps_{Q-1}))
   \delta(\xi_{n+m+1}-\xi_{n+1}+\xi_n-\xi_0)
   d\bs d\bxi.
   \end{array}
\end{equation}
Note that neither $n$ nor $n+m+1$ belong to $A'(\mg)$. For each $k\in
A'(\mg)$, we integrate in $s_k$ and obtain using \eqref{eq:bdtime}
that
\begin{equation}
  \label{eq:Vepsmnbd1}
  \begin{array}{l}
  |V_\eps^{n,m}(t,\xi_0,\xi_{n+m+1})| \leq \dsum_{\mg\in\mG_c}
  \dint \dprod_{k\not\in A'(\mg)} e^{-s_k \xi_k^\m}
      |\hat u_0(\xi_n) \bar {\hat u}_0(\xi_{n+1})| \\
 \qquad 
   \dprod_{k\in A'(\mg)}
   \eps^{-2\alpha} \Big(\dfrac{\eps^\m}{\xi_k^\m}\wedge t\Big)
   \hat R(\xi_k-\eps\xi^\eps_{k-1})
   \delta(\frac{\xi_{k}}\eps-\xi^\eps_{k-1}+\xi_{l(k)}-\xi^\eps_{l(k)-1})
   \\\qquad
    \eps^{d-2\alpha} \hat R(\eps(\xi_Q-\xi^\eps_{Q-1}))
   \delta(\xi_{n+m+1}-\xi_{n+1}+\xi_n-\xi_0)
   d\tilde\bs d\bxi.
   \end{array}
\end{equation}
By assumption on $\hat R(\xi)$, we know the existence of a constant
$\hat R_\infty$ such that 
\begin{equation} \label{eq:unifbdhatR}
   \eps^{d-2\alpha} \hat R(\eps(\xi_Q-\xi^\eps_{Q-1})) \leq 
   \eps^{d-2\alpha} \hat R_\infty.
\end{equation}
This is where the factor $\eps^{d-2\alpha}$ arises. We need however to
ensure that the integral in $\xi_Q$ is well-defined. We have two
possible scenarios: either $Q=n$ or $n\in B_0$. When $Q=n$, the
integration in $\xi_Q$ is an integration in $\xi_n$ for which we use
$\hat u_0(\xi_n)$. When $n\in B_0$, we thus have $n=l(k_0)$ for some
$k_0$ and we replace the delta function involving $\xi_n$ by a delta
function involving $\xi_Q$ given equivalently by
\begin{equation} \label{eq:deltause}
  \delta(\xi_Q-\xi_{Q-1} + \xi_0-\xi_n + \dsum_{m=1}^M\xi_{q_m}-\xi_{q_m-1}).
\end{equation}
In either scenario, we can integrate in the variable $\xi_Q$ without
using the term $\hat R(\eps(\xi_Q-\xi_{Q-1}))$.  We use the inequality
\begin{equation}\label{eq:ineq}
  |\hat u_0(\xi_n) \bar {\hat u}_0(\xi_{n+1})|
\leq \dfrac12\Big(|\hat u_0(\xi_n) |^2 + |\hat u_0(\xi_n-\xi_0+\xi_{n+m+1})|^2 
  \Big),
\end{equation}
to obtain the bound
\begin{equation}
  \label{eq:Vepsmnbd2}
  |V_\eps^{n,m}(t,\xi_0,\xi_{n+m+1})| \leq  \eps^{d-2\alpha} \hat R_\infty
  \dsum_{\mg\in\mG_c}
  \Big(\dint d\tilde s\Big) \rho_f^{\ban-1} \|\hat u_0\|^2.
\end{equation}
The bound is uniform in $\xi_0$ and $\xi_{n+m+1}$. Using
\eqref{eq:boundgraphs} and \eqref{eq:bdVeps1}, we obtain
\begin{equation}
  \label{eq:Vepsmnbd3}
  |V_\eps^{n,m}(t,\xi_0,\xi_{n+m+1})| \leq 
   \eps^{d-2\alpha} \rr^{n+m}\|\hat u_0\|^2.
\end{equation}
After summation in $n,m\in\Nm$, we thus find that 
\begin{equation}
  \label{eq:Xeps2}
  \E \{ (X_\eps - \E\{X_\eps\})^2\} \leq \dfrac{C}{\rr^2}
   \eps^{d-2\alpha} \|\hat u_0\|^2 \|\hat M\|_1^2.
\end{equation}
Similarly, by setting $\xi_{n+m+1}=\xi_0$, we find that 
\begin{equation}
  \label{eq:locenergy}
  \Big|\E\Big\{ \dint_{\Rm^d} |\hat u_\eps|^2(t,\xi) \varphi(\xi) d\xi
  \Big\} - \dint_{\Rm^d} 
   |\E\{\hat u_\eps(t,\xi)\}|^2 \varphi(\xi) d\xi \Big|
  \leq \dfrac{C}{\rr^2}\eps^{d-2\alpha} \|\hat u_0\|^2 \|\varphi\|_1,
\end{equation}
for any test function $\varphi\in L^1(\Rm^d)$.  This local energy
estimate is to be compared with the global estimate obtained in
\eqref{eq:boundI}.

\paragraph{Analysis of the leading crossing terms in $X_\eps$.}
The preceding estimate on $X_\eps$ may be refined as only the crossing
graphs in $\mG_{cs}$ have contributions of order $\eps^{d-2\alpha}$.
We return to the bound \eqref{eq:Vepsmnbd1} and obtain that
\begin{equation}
  \label{eq:Vepsmnbd4}
  \begin{array}{l}
  |V_\eps^{n,m}(t,\xi_0,\xi_{n+m+1})| \leq 
  \eps^{d-2\alpha} \hat R_\infty
  \dsum_{\mg\in\mG_c}
  \dint \dprod_{k\not\in A'(\mg)} e^{-s_k(\xi_k)^\m}
      |\hat u_0(\xi_n) \bar {\hat u}_0(\xi_{n+1})| \\
 \qquad 
   \dprod_{k\in A'(\mg)}
    \eps^{-2\alpha} \Big(\dfrac{\eps^\m}{\xi_k^\m}\wedge t\Big)
    \hat R(\xi_k-\eps\xi^\eps_{k-1})
   \delta(\frac{\xi_{k}}\eps-\xi^\eps_{k-1}+\xi_{l(k)}-\xi^\eps_{l(k)-1})
   \\\qquad
   \delta(\xi_{n+m+1}-\xi_{n+1}+\xi_n-\xi_0)
   d\tilde\bs d\bxi.
   \end{array}
\end{equation}
The $\ban+3$ variables in time left are $s_0$, $s_{n+1}$, $s_Q$,
$s_{l(Q)}$, and the $\ban-1$ variables $s_{l(A'(\mg))}$.

Let $\mg\in\mG_{c}$.  Let us assume that for some $k$ such that
$(\xi_k,\xi_{l(k)})$ is not a crossing pair, we have $l(k)-1>k$,
i.e., $\mg\in\mG_{ncs}$.  The non-crossing pairs are not affected by
the possible change of a delta function involving $\xi_n$ to a delta
function involving $\xi_Q$.  We may then integrate in the variable
$s_{l(k)}$ and obtain the bound for the integral
\begin{displaymath}
  \begin{array}{ll}
  \eps^{d-2\alpha} \hat R_\infty
  \dint  d\tilde\bs d\bxi  |\hat u_0(\xi_n) \bar {\hat u}_0(\xi_{n+1})| 
   \delta(\xi_{n+m+1}-\xi_{n+1}+\xi_n-\xi_0)  \\[3mm]
   \dprod_{k\in A'(\mg)}
    \Big(\dfrac{\eps^\m}{\xi_k^\m}\wedge t\Big) 
    \Big( \dfrac{\eps^\m}{|\xi_k-\eps\xi_{k-1}^\eps
     - \eps\xi_{l(k)-1}^\eps|^\m} \wedge t\Big)
  \hat R(\xi_k-\eps\xi^\eps_{k-1})
   \delta(\frac{\xi_{k}}\eps-\xi^\eps_{k-1}+\xi_{l(k)}-\xi^\eps_{l(k)-1})\\
   \leq \eps^\beta \eps^{d-2\alpha}  \Big(\dint d\tilde\bs \Big) 
   \hat R_\infty \rho_f^{\ban-1}\|\hat u_0\|^2,
   \end{array}
\end{displaymath}
thanks to lemma \ref{lem:bound} below.  The summation over all
graphs in $\mG_{ncs}$ of any quantity derived from
$V_\eps^{n,m}(t,\xi_0,\xi_{n+m+1})$ is therefore $\eps^\beta$ smaller
than the corresponding sum over all graphs in $\mG_c$.  We thus see
that any non-crossing pair has to be of the form $l(k)-1=k$, i.e., a
simple pair, in order for the graph to correspond to a contribution of
order $\eps^{d-2\alpha}$.

Let us consider the graphs composed of crossings and simple pairs.  We
may delete the simple pairs from the graph since they contribute
integrals of order $O(1)$ thanks to lemma \ref{lem:bound} below
and assume that the graph is composed of crossings only, thus with
$n=m$ and $Q=n$ after deletion of the simple pairs. Let us consider
$k<n$ with $l(k)\geq n+1$ so that the delta function
\begin{displaymath}
    \delta(\dfrac{\xi_{k}}\eps - \xi_{k-1}^\eps 
     + \xi_{l(k)}-\xi_{l(k)-1})
\end{displaymath}
is present in the integral defining $V^{n,m}_\eps$.  We find for the
same reason as above that the contribution of the corresponding graph
is of order $\eps^{d-2\alpha}\eps^\beta$ by integration in the
variable $s_{l(k)}$.  As a consequence, the only graph composed
exclusively of crossing pairs that generates a contribution of order
$\eps^{d-2\alpha}$ is the graph with $n=m=1$.  This concludes our
proof that the contribution of order $\eps^{d-2\alpha}$ in
$V^{n,m}_\eps$ is given by the $nm$ graphs in $\mG_{cs}$ when both $n$
and $m$ are even numbers (otherwise, $\mG_{cs}$ is empty). All other
graphs in $\mG_c$ provide a contribution of order $\eps^\beta$ smaller
than what we obtained in \eqref{eq:Vepsmnbd3}.
In other words, let us define 
\begin{equation}
  \label{eq:Vepsmn}
  \begin{array}{l}
 V_{\eps,s}^{n,m}(t,\xi_0,\xi_{n+m+1}) =\dsum_{\mg\in\mG_{cs}} 
  \dint \prod_{k=0}^{n+m+1} e^{-s_k\xi_k^\m}
      \hat u_0(\xi_n) \bar {\hat u}_0(\xi_{n+1}) \\
 \qquad  \dprod_{k\in A_0(\mg)}
   \eps^{d-2\alpha} \hat R(\eps(\xi_k-\xi_{k-1}))
  \delta(\xi_{k}-\xi_{k-1}+\xi_{l(k)}-\xi_{l(k)-1})
   d\bs d\bxi.
  \end{array}
\end{equation}
We have found that 
\begin{equation}
  \label{eq:bounddeltaV}
  |V_{\eps}^{n,m}(t,\xi_0,\xi_{n+m+1})-V_{\eps,s}^{n,m}(t,\xi_0,\xi_{n+m+1})|
  \lesssim \eps^{d-2\alpha+\beta}  \rr^{n+m}\|\hat u_0\|^2.
\end{equation}

\subsection{Analysis of non-crossing graphs}
\label{sub:noncrossing}
We now apply the estimates obtained in the preceding section to the
analysis of the moments $U_\eps^n(t)$ defined in \eqref{eq:Uepsn} and
given more explicitly in \eqref{eq:Uepsm}. Our objective is to show
that only the simple graph $\mg$ contributes a term of order $O(1)$ in
\eqref{eq:Uepsm} whereas all other graphs in $\mG_{ns}$ contribute
(summable in $n$) terms of order $O(\eps^\beta)$. Note that $n=2\ban$,
for otherwise, $U_\eps^n(t)=0$. We recall that the simple graph is
defined by $l(k)=k+1$. We thus define the simple graph contribution as
 \begin{equation}
  \label{eq:Uepsns}
  \begin{array}{rcl}
  U_{\eps,s}^n(t,\xi_0) &=& {\mathcal U}^n_\eps(t,\xi_0) \hat u_0(\xi_0)
  \\ {\mathcal U}^n_\eps(t,\xi_0)&=&
    \dint \prod_{k=0}^n e^{-s_k\xi_k^\m} \prod_{k=0}^{\ban-1}
     \eps^{d-2\alpha} \hat R(\eps(\xi_{2k+1}-\xi_{2k}))
   \delta(\xi_{2(k+1)}-\xi_{2k}) d\bs d\bxi,
   \end{array}
\end{equation}
and 
\begin{equation}
  \label{eq:Uepss}
  U_{\eps,s}(t,\xi_0)=\dsum_{n\in\Nm} U_{\eps,s}^n(t,\xi_0)
   := {\mathcal U}_\eps(t,\xi_0) \hat u_0(\xi_0).
\end{equation}
For all $k\in A_0$, we perform the change of variables
$\xi_k\to\frac{\xi_k}\eps$ and (re-)define as before
\begin{equation}
  \label{eq:xieps3}
  \xi_k^\eps = \left\{
    \begin{array}{ll}
          \xi_k & k\not\in A_0 \\
          \frac{\xi_k}\eps & k\in A_0.
    \end{array}\right.
\end{equation}
This gives
\begin{equation}
  \label{eq:Uepsn2}
  \begin{array}{l}
  U_\eps^n(t,\xi_0) =  \hat u_0(\xi_0) 
   \dsum_{\mg\in\mG} \dint \prod_{k=0}^n e^{-s_k(\xi_k^\eps)^\m}
   \\\qquad \dprod_{k\in A_0(\mg)}
   \eps^{-2\alpha} \hat R(\xi_k-\eps\xi^\eps_{k-1})
  \delta(\dfrac{\xi_{k}}\eps-\xi^\eps_{k-1}+\xi_{l(k)}-\xi^\eps_{l(k)-1}) 
   d\bs d\bxi.
  \end{array}
\end{equation}
Assuming that $l(k)-1>k$ for one of the pairings, we obtain as in the
analysis leading to \eqref{eq:bounddeltaV} the following bound for the
corresponding graph:
\begin{displaymath}
  \begin{array}{l}
   |\hat u_0(\xi_0)| \dint  d\tilde\bs d\bxi  \dprod_{k\in A_0(\mg)}
    \eps^{-2\alpha}\Big(\dfrac{\eps^\m}{\xi_k^\m}\wedge t\Big)
     \Big( \dfrac{\eps^\m}{|\xi_k-\eps\xi_{k-1}^\eps
     - \eps\xi_{l(k)-1}^\eps|^\m} \wedge t\Big)\\\qquad\qquad
  \hat R(\xi_k-\eps\xi^\eps_{k-1})
   \delta(\dfrac{\xi_{k}}\eps-\xi^\eps_{k-1}+\xi_{l(k)}-\xi^\eps_{l(k)-1})
   \\[2mm]
   \leq \eps^{\beta}   \Big(\dint d\tilde\bs \Big) 
   \rho_f^{\ban}|\hat u_0(\xi_0)|.
   \end{array}
\end{displaymath}
This shows that 
\begin{equation}
  \label{eq:limUepsn}
  |U_\eps^n(t,\xi_0) - U_{\eps,s}^n(t,\xi_0)|
  \leq |\hat u_0(\xi_0)|  \eps^{\beta} \rr^{n},
\end{equation}
so that 
\begin{equation}
  \label{eq:limavueps}
  |\E\{\hat u_\eps\}(t,\xi)- U_{\eps,s}(t,\xi)| \lesssim \dfrac{1}{\rr}
  \eps^\beta |\hat u_0(\xi)| ,
\end{equation}
at least for sufficiently small times $t\in(0,T)$ such that
$4\rho_fT<1$. It remains to analyze the limit of $U_{\eps,s}(t,\xi)$
to obtain the limiting behavior of $X_\eps$ and $I_{\eps,\varphi}$.
This analysis is carried out in the next section. Another application
of lemma \ref{lem:bound} shows that $U_{\eps,s}(t,\xi)$ is square
integrable and that its $L^2(\Rm^d)$ norm is bounded by $\|\hat
u_0\|$. In other words, we have constructed a weak solution $\hat
u_\eps(t)\in L^2(\Omega\times \Rm^d)$ to \eqref{eq:parabFD} since the
series \eqref{eq:exp} converges uniformly in $L^2(\Omega\times \Rm^d)$
for sufficiently small times $t\in(0,T)$ such that $4\rho_fT<1$.  

Collecting the results obtained in \eqref{eq:boundI} and
\eqref{eq:limavueps}, we have shown that
\begin{displaymath}
  \begin{array}{rcl}
   \|(\hat u_\eps-U_{\eps,s})(t)\|_{L^2(\Omega\times \Rm^d)} &\lesssim& 
  \eps^{\frac{\beta}2} \|\hat u_0\|_{L^2(\Rm^d)},
  \end{array}
\end{displaymath}
where $U_{\eps,s}$ is the deterministic term given in
\eqref{eq:Uepss}. The analysis of $U_{\eps,s}$ and that of $X_\eps$ is
postponed to section \ref{sec:limits}, after we state and prove lemma
\ref{lem:bound}, which allows us to analyze the contributions of
the different graphs.

\begin{lemma}
  \label{lem:bound}
  Let us assume that $\hat R$ is bounded by a smooth radially
  symmetric, decreasing function $f(r)$. We also assume that $f(r)\leq
  \tau_f r^{-\n}$ for some $0\leq\n<d-\m$ in dimension $d>\m$ and
  $\n=0$ when $d\leq\m$. Then we obtain the following estimates.\\ For
  $d>\m$, we have
  \begin{displaymath}
    \dint \dfrac{1}{|\xi_k|^\m} \hat R(\xi_k-y) d\xi_k \leq \rho_f
     := c_d \dint_0^\infty \dfrac{1}{|\xi|^\m} f(|\xi|)|\xi|^{d-1}d|\xi|
      \,\,\vee \,\, \tau_f,
  \end{displaymath}
  uniformly in $y\in\Rm^d$, where $c_d=|S^{d-1}|$ and $a\vee
  b=max(a,b)$. Moreover,
  \begin{displaymath}
    \dint \dfrac{1}{|\xi_k|^\m} \hat R(\xi_k-y)
   \Big( \dfrac{\eps^\m}{|\xi_k-z|^\m}\wedge t\Big) d\xi_k
   \lesssim \,\rho_f \,
    \left\{
    \begin{array}{ll}
       \eps^{\m-\n} \quad &d> 2\m-\n \\
       \eps^{\m-\n}|\ln\eps| \quad &d=2\m-\n\\
       \eps^{d-\m-\n}  \quad & \m<d<2\m-\n.
    \end{array} 
    \right.
  \end{displaymath}
  For $d=\m$, we define $\rho_f = c_d f(0)$ and  have
  \begin{displaymath}
    \dint \Big( \dfrac{\eps^\m}{|\xi_k-z|^\m}\wedge t\Big)^l\hat R(\xi_k-y) 
    d\xi_k \lesssim \,\rho_f \,\left\{
      \begin{array}{ll}
        \eps^\m|\ln\eps| & l=1 \\
        \eps^\m & l=2.
      \end{array}\right.
  \end{displaymath}
  For $d< \m$, we have
  \begin{displaymath}
    \dint \Big( \dfrac{\eps^\m}{|\xi_k-z|^\m}\wedge t\Big)^l\hat R(\xi_k-y) 
    d\xi_k  \lesssim \eps^d,\quad l\geq1.
  \end{displaymath}
\end{lemma}
\begin{proof}
  Once $\hat R$ is bounded above by a decreasing, radially symmetric,
  function $f(r)$, the above integrals are maximal when $y=z=0$ thanks
  to lemma \ref{lem:decay} below since $|\xi|^{-\m}$ and
  $(\eps^\m|\xi|^{-\m}\wedge t)$ are radially symmetric and decreasing.
  The first bound is then obvious and defines $\rho_f$. The second
  bound is obvious in dimension $d>2\m$ since $|\xi_k|^{-2\m}$ is
  integrable.

  All the bounds in the lemma are thus obtained from a bound for
  \begin{displaymath}
    \dint_0^\infty \Big( \dfrac{\eps^\m}{r^\m}\wedge t\Big)^l 
   r^{d-1} f(r) dr.
  \end{displaymath}
  We obtain that the above integral restricted to $r\in (1,\infty)$ is
  bounded by a constant times $\eps^{\m l} \rho_f$ for $d\geq\m$ and
  by a constant times $\eps^{\m l}$ for $d<\m$. It thus remains to bound
  the integral on $r\in(0,1)$, which is equal to
  \begin{displaymath}
    \dint_0^{\eps t^{-\frac1{\m}}} t^l r^{d-1} f(r) dr
    + \dint_{\eps t^{-\frac1{\m}}}^1 \dfrac{\eps^{l\m}}{r^{l\m}}r^{d-1} f(r)dr.
  \end{displaymath}
  Replacing $f(r)$ by $\tau_f r^{-\n}$, we find that the first
  integral is bounded by a constant times $\eps^{d-\n}$ and the second
  integral by a constant times $\eps^{d-\n} \vee \eps^{l\m}$ when
  $d-\n-l\m\not=0$ and $\eps^{2\m} |\ln\eps|$ when $d=2\m-\n$. It
  remains to divide through by $\eps^\m$ when $l=2$ to obtain the
  desired results.
\end{proof}
\begin{lemma}
  \label{lem:decay}
  Let $f$, $g$, and $h$ be non negative, bounded, integrable, and
  radially symmetric functions on $\Rm^d$ that are decreasing as a
  function of radius.  Then the integral
  \begin{equation}
    \label{eq:intab}
    I_{\zeta,\tau} = \dint_{\Rm^d} f(\xi-\zeta)g(\xi-\tau)h(\xi) d\xi,
  \end{equation}
  which is well defined, is maximal at $\zeta=\tau=0$.
\end{lemma}
\begin{proof}
  In a first step, we rotate $\zeta$ to align it with $\tau$. The
  first claim is that the integral cannot increase while doing so.
  Then we send $\zeta$ and $\tau$ to $0$. The second claim is that the
  integral again does not increase. 
  
  We assume that the functions $f$, $g$, and $h$ are smooth and obtain
  the result in the general case by density. We choose a system of
  coordinates so that $\tau=|\tau|e_1$, where $(e_1,\ldots,e_d)$ is an
  orthonormal basis of $\Rm^d$, and $\zeta=|\zeta|\hat\theta$ with
  $\hat \theta=(\cos\theta,\sin\theta,0,\ldots,0)$. Without loss of
  generality, we may assume that $\theta\in(0,\pi)$. Then
  $I_{\zeta,\tau}$ may be recast as $I_\theta$ and we find that
  \begin{displaymath}
    I_\theta = \dint_0^\infty |\xi|^{d-1} h(|\xi|) J_\theta(|\xi|) d|\xi|,
  \end{displaymath}
  where we denote $h(|\xi|)\equiv h(\xi)$ with the same convention for
  $f$ and $g$ and define
  \begin{displaymath}
    J_\theta(|\xi|) = \dint_{S^{d-1}} f(|\xi|\psi-\zeta)g(|\xi|\psi-\tau)
   d\psi. 
  \end{displaymath}
  It is sufficient to show that $\partial_\theta J_\theta \leq0$. We
  find
  \begin{displaymath}
    \partial_\theta J_\theta = \dint_{S^{d-1}} -\hat\theta^\perp\cdot
  \nabla f(|\xi|\psi-\zeta) g(|\xi|\psi-\tau) d\psi,
  \end{displaymath}
  with $\hat\theta^\perp=(-\sin\theta,\cos\theta,0,\ldots,0)$.
  We decompose the sphere as $\psi=(\psi\cdot\hat\theta,\tilde\psi)$
  and find, for some positive weight $w(\mu)$ that
  \begin{displaymath}
    \begin{array}{rcl}
    \partial_\theta J_\theta &=& \dint_{-1}^1 d(\psi\cdot\hat\theta)
   (-f')(||\xi|\psi-\zeta|) w(\psi\cdot\hat\theta)
   \dint_{S^{d-2}} (\hat\theta^\perp\cdot\tilde\psi) g(|\xi|\psi-\tau)  
   d\tilde\psi.
    \end{array}
  \end{displaymath}
  We now observe that
  \begin{displaymath}
    \begin{array}{rcl}
    && \dint_{S^{d-2}} (\hat\theta^\perp\cdot\tilde\psi) g(|\xi|\psi-\tau)  
   d\tilde\psi\\[3mm]
   &=&\dint_{\hat\theta^\perp\cdot\tilde\psi>0} 
     (\hat\theta^\perp\cdot \tilde\psi )\big(
     g(||\xi|(\hat\theta\cdot\psi\hat\theta+\tilde\psi)-\tau|)
    - g(||\xi|(\hat\theta\cdot\psi\hat\theta-\tilde\psi)-\tau|) \big)
   d\tilde\psi\,\leq\,0,
    \end{array}
  \end{displaymath}
  as
  $||\xi|(\hat\theta\cdot\psi\hat\theta+\tilde\psi)-\tau|\leq||\xi|(\hat\theta\cdot\psi\hat\theta-\tilde\psi)-\tau|$
  by construction. Indeed, we find that
  $||\xi|(\hat\theta\cdot\psi\hat\theta\pm\tilde\psi)-\tau|^2-|\xi|^2-|\tau|^2+2|\tau||\xi|\hat\theta\cdot\psi\hat\theta\cdot\tau=\pm2|\tau||\xi|\tilde\psi\cdot\tau=\pm2|\tau||\xi|\hat\theta^\perp\cdot\tau$
  whereas $\hat\theta^\perp\cdot\tau=-\sin\theta|\tau|<0$ by
  construction. This shows that
  $|\xi|(\hat\theta\cdot\psi\hat\theta+\tilde\psi)$ is closer to
  $\tau$ than $|\xi|(\hat\theta\cdot\psi\hat\theta-\tilde\psi)$ is,
  and since $g(r)$ is decreasing, that $\partial_\theta
  J_\theta\leq0$. This concludes the proof of the first claim.
  
  If $\beta=0$ or $\tau=0$, we set $b=0$ below. Otherwise, we may
  assume without loss of generality that $\tau=-b\zeta$ for some
  $b>1$.  We still define $\zeta=|\zeta|\hat\theta$. We now define the
  integral
  \begin{math}
    I_a = I_{a\zeta,b\zeta},\,0\leq a\leq1,
  \end{math}
  and compute
  \begin{displaymath}
    \partial_a I_a = \dint_{\Rm^d} -\zeta\cdot\nabla f(\xi-a\zeta)
   g(\xi+b\zeta)h(\xi) d\xi=
   \dint_{\Rm^d} -\zeta\cdot\nabla f(\xi) g(\xi+(b-a)\zeta)h(\xi+a\zeta)
   d\xi. 
  \end{displaymath}
  Define $l(\xi,\zeta)=g(\xi+(b-a)\zeta)h(\xi+a\zeta)$. Then because
  $f$ is radially symmetric, we have
  \begin{displaymath}
    \partial_a I_a = \dint_0^\infty m(|\xi|) |\xi|^{d-1} d |\xi|,\,\,\quad
   m(|\xi|) = -f'(|\xi|) \dint_{S^{d-1}} \hat\theta\cdot\psi\,
        l(|\xi|\psi,\zeta) d\psi.
  \end{displaymath}
  We recast 
  \begin{displaymath}
    m(|\xi|) = -f'(|\xi|) \dint_{\hat\theta\cdot\psi>0}  (\hat\theta\cdot\psi)
    \big( l(|\xi|\psi,\zeta) - l (-|\xi|\psi,\zeta)\big) d\psi
    \leq0,
  \end{displaymath}
  since $\big||\xi|\psi+\gamma\zeta\big|\geq
  \big|-|\xi|\psi+\gamma\zeta\big|$ by construction for all $\gamma>0$
  and thus for $\gamma=a$ and $\gamma=b-a$. This shows that
  $\partial_\alpha I_\alpha\leq0$ and concludes the proof of the
  second claim.
\end{proof}
\section{Homogenized limit and Gaussian fluctuations}
\label{sec:limits}

In this section, we conclude the proof of theorems
\ref{thm:convergence} and \ref{thm:fluct}.

\subsection{Homogenization theory for $u_\eps$}
\label{sec:convergence}

We come back to the analysis of $U_{\eps,s}(t,\xi)$ defined in
\eqref{eq:Uepsns}. Since only the simple graph is retained in the
definition of mean field solution $U_{\eps,s}(t,\xi)$, the equation it
satisfies may be obtained from that for $\hat u_\eps$ by simply
assuming the mean field approximation $\E\{\hat q_\eps \hat q_\eps
\hat u_\eps\} \sim \E\{\hat q_\eps \hat q_\eps\}\E\{\hat u_\eps\}$
since the Duhamel expansions then agree.  As a consequence, we find
that $\Ues$ is the solution to the following integral equation
\begin{equation}
  \label{eq:intUes}
  \begin{array}{l}
 \Ues(t,\xi)= e^{-t\xi^\m} \hat u_0(\xi) \\\quad + \dint_0^te^{-\xi^\m s}
   \dint_0^{t-s} e^{-\xi_1^\m s_1} \dint\eps^{d-2\alpha}\hat R(\eps(\xi_1-\xi))
   \Ues(t-s-s_1,\xi) d\xi_1 dsds_1\\
  =  e^{-t\xi^\m} \hat u_0(\xi) + \dint_0^t \dint_0^v
    e^{-\xi^\m (v-s_1)} e^{-\xi_1^\m s_1}
   \eps^{d-2\alpha}\dint\hat R(\eps(\xi_1-\xi))
   \Ues(t-v,\xi) d\xi_1 ds_1 dv \\
  =  e^{-t\xi^\m} \hat u_0(\xi) + \eps^{\m-2\alpha}
   \dint_0^t \dint_0^{\frac v{\eps^\m}} e^{-\xi^\m (v-\eps^\m s_1)} 
    e^{-\xi_1^\m s_1} \dint\hat R(\xi_1-\eps\xi) d\xi_1 ds_1
   \Ues(t-v,\xi) dv \\
    := e^{-t\xi^\m} \hat u_0(\xi) + A_\eps \Ues (t,\xi).
  \end{array}
\end{equation}
The last integral results from the change of variables
$\eps\xi_1\to\xi_1$ and $s_1\eps^{-\m}\to s_1$.  It remains to analyze
the convergence properties of the solution to the latter integral
equation. Note that $\xi$ acts as a parameter in that equation. Let us
decompose
\begin{equation}
  \label{eq:decompAeps}
  A_\eps U(t,\xi) = \rho_\eps \dint_0^t e^{-\xi^\m v} U(t-v,\xi) dv
   + E_\eps U(t,\xi),
\end{equation}
with $\rho_\eps=\int_{\Rm^d} \frac{\hat R(\xi_1-\eps\xi)}{\xi_1^\m}
d\xi_1$ when $d>\m$ and $\rho_\eps=c_d\hat R(\eps\xi)$ when $d=\m$.
Then we have
\begin{lemma}
   \label{lem:Eeps}
   Let $\xi\in\Rm^d$ and $f(r)$ as in lemma \ref{lem:bound}.  Then the
   operator $E_\eps$ defined above in \eqref{eq:decompAeps} is bounded
   in the Banach space of continuous functions on $(0,T)$.  Moreover,
   we have
  \begin{equation}
    \label{eq:boundCeps}
    \|E_\eps\|_{\mathcal L(\mathcal C(0,T))} \lesssim \eps^{\beta-\n}.
  \end{equation}
\end{lemma}
\begin{proof}
  We start with the case $d>\m$ so that and $\eps^{\m-2\alpha}=1$.
  Note that $\n$ in lemma \ref{lem:bound} is defined such that
  $d>\m-\n$ as well.  With $B_\eps=A_\eps-E_\eps$ in
  \eqref{eq:decompAeps}, we find that
  \begin{displaymath}
     B_\eps \Ues(t,\xi) = \dint_0^t e^{-\xi^\m v} \dint_0^\infty 
   \dint e^{-\xi_1^\m s_1}
   \hat R(\xi_1-\eps\xi)  d\xi_1ds_1 \Ues(t-v,\xi) dv.
  \end{displaymath}
  The remainder $E_\eps$ is then given by
  \begin{displaymath}
  \begin{array}{ll}
  E_\eps \Ues(t,\xi) &=  \dint_0^t \dint_0^{\frac v{\eps^\m}} \dint
    e^{-\xi^\m v} (e^{\eps^\m\xi^\m s_1}-1) 
    e^{-\xi_1^\m s_1} \hat R(\xi_1-\eps\xi) d\xi_1 ds_1
   \Ues(t-v,\xi) dv\\
  & - \dint_0^t \dint_{\frac v{\eps^\m}}^\infty  \dint
     e^{-\xi^\m v} 
    e^{-\xi_1^\m s_1} \hat R(\xi_1-\eps\xi) d\xi_1 ds_1
   \Ues(t-v,\xi) dv.
  \end{array}
  \end{displaymath}
  The continuity of $E_\eps \Ues(t,\xi)$ in time is clear when
  $\Ues(t,\xi)$ is continuous in time. Without loss of generality, we
  assume that $\Ues(\cdot,\xi)$ is bounded by $1$ in the uniform norm.
  We decompose the integral in the $s_1$ variable in the first term of
  the definition of $E_\eps$ into two integrals on $0\leq s_1\leq
  \frac{v}{2\eps^\m}$ and $\frac{v}{2\eps^\m}\leq s_1\leq \frac
  v{\eps^\m}$. Because $e^{-\xi^\m v} (e^{\eps^\m\xi^\m s_1}-1)\leq1$,
  the second integral is estimated as
  \begin{displaymath}
  \begin{array}{l}
  \dint_0^t \dint_{\frac v{2\eps^\m}}^{\frac v{\eps^\m}} \dint
    e^{-\xi^\m v} (e^{\eps^\m\xi^\m s_1}-1) 
    e^{-\xi_1^\m s_1} \hat R(\xi_1-\eps\xi) d\xi_1 ds_1 dv \\[4mm]
  \leq \dint_0^t  \dint \dfrac{1}{\xi_1^\m}e^{-\xi_1^\m\frac{v}{2\eps^\m}}
   \hat R(\xi_1-\eps\xi) d\xi_1  dv 
  \leq \dint \dfrac{2}{\xi_1^\m}\Big(\dfrac{\eps^\m}{\xi_1^\m}\wedge t\Big)
   \hat R(\xi_1-\eps\xi) d\xi_1 \lesssim \eps^{\beta-\n} \rho_f,
  \end{array}
  \end{displaymath}
  thanks to lemma \ref{lem:bound}. The above bound is uniform in
  $\xi$. The last integral defining $E_\eps$ on the interval $s_1\geq
  \frac{v}{\eps^\m}$ is treated in the exact same way and also
  provides a contribution of order $O(\eps^{\beta-\n})$.  
  
  The final contribution involves the integration over the interval
  $0\leq s_1\leq\frac v{2\eps^\m}$. Using $e^{-\xi^\m v}
  (e^{\eps^\m\xi^\m s_1}-1)\leq \eps^\m\xi^\m s_1 e^{-\frac{\xi^\m
      v}2}$ on that interval, it is bounded by
  \begin{displaymath}
    \begin{array}{l}
    I_3 :=\dint_0^t \dint_0^{\frac{v}{2\eps^\m}} \dint_{\Rm^d}
    \eps^\m \xi^\m s_1 e^{-\frac{\xi^\m v}2}
   e^{-\xi_1^\m s_1} \hat R(\xi_1-\eps\xi) d\xi_1 ds_1 dv \\
   \leq \dfrac{2\eps^\m \xi^\m }{\xi^\m}\big(1-e^{-\frac{\xi^\m t}2}\big)
    \dint_0^{\frac{t}{2\eps^\m}}  s_1\dint_{\Rm^d}   e^{-\xi_1^\m s_1}
     \hat R(\xi_1-\eps\xi) d\xi_1 ds_1,
    \end{array}
  \end{displaymath}
  by switching the variables $0\leq s\leq \frac{v}{2\eps^\m} \leq
  \frac{t}{2\eps^\m}$. Using lemma \ref{lem:decay}, we may replace
  $\hat R(\xi_1-\eps\xi)$ by $\hat R(\xi_1)$ in the above expression.
  This shows that
  \begin{displaymath}
    I_3 \leq 2\eps^\m  \dint_{\Rm^d} \dint_0^{\frac{t}{2\eps^\m}}s_1 
   e^{-\xi_1^\m s_1}ds_1 \hat R(\xi_1) d\xi_1.
  \end{displaymath}
  We observe that
  \begin{displaymath}
    \dint_0^{\tau} s_1 
   e^{-\xi_1^\m s_1}ds_1 \lesssim \dfrac{1}{\xi_1^{2\m}} \wedge  \tau^2,   
  \end{displaymath}
  so that 
  \begin{displaymath}
    I_3 \lesssim \eps^\m\dint_0^\infty f(r) r^{d-1} 
    \big(r^{-2\m}\wedge \tau^2\big)dr,\qquad \tau = \dfrac{t}{2\eps^\m}\vee 1.
  \end{displaymath}
  The integral over $(1,\infty)$ is bounded by $\eps^\m\rho_f$. Using
  the assumption that $f(r)\lesssim r^{-\n}$, we obtain that the
  integral over $(0,1)$ is bounded by a constant times
  \begin{displaymath}
    \dint_0^{\tau^{-\frac 1\m}} r^{d-1-\n} dr
    + \dint_{\tau^{-\frac 1\m}}^1 r^{d-1-\n-2\m} dr \lesssim
    \tau^{2-\frac{d-\n}{\m}} \vee 1,
  \end{displaymath}
  when $d-\n-2\m\not=0$ and $|\ln\tau|$ when $d=\n+2\m$. Since $\tau$
  is bounded by a constant times $\eps^{-\m}$, this shows that $I_3$
  is bounded by $\eps^{d-\m-\n}$ when $d-\n-2\m\not=0$ and
  $\eps^d|\ln\eps|$ when $d=\n+2\m$. This concludes the proof when
  $d>\m-\n$.

  \medskip
  
  We now consider the proof when $d=\m$ with $\n=0$. Then,
  $\eps^{\m-2\alpha}=\frac{1}{|\ln\eps|}$. The leading term is given
  by $U_{\eps,s}$, which solves the integral equation:
  \begin{equation}
  \label{eq:intUes2s}
  \begin{array}{l}
 \Ues(t,\xi)= e^{-t\xi^\m} \hat u_0(\xi) \\\quad + \dint_0^te^{-\xi^\m s}
   \dint_0^{t-s} e^{-\xi_1^\m s_1} \dint \dfrac{1}{|\ln\eps|}
    \hat R(\eps(\xi_1-\xi))
   \Ues(t-s-s_1,\xi) d\xi_1 dsds_1\\
  =  e^{-t\xi^\m} \hat u_0(\xi) + \dfrac{1}{|\ln\eps|}
    \dint_0^t \dint_0^{\frac v{\eps^\m}}
    e^{-\xi^\m (v-\eps^\m s_1)} 
    e^{-\xi_1^\m s_1} \hat R(\xi_1-\eps\xi) d\xi_1 ds_1
   \Ues(t-v,\xi) dv \\
    = e^{-t\xi^\m} \hat u_0(\xi) + A_\eps \Ues (t,\xi),\qquad
   A_\eps = B_\eps+E_\eps.
  \end{array}
  \end{equation}
  Here we have defined
  \begin{displaymath}
  B_\eps U (t,\xi) = \rho_\eps \dint_0^t e^{-\xi^\m v} U(t-s,\xi)ds,\quad
   \rho_\eps = c_d \hat R(\eps\xi),
  \end{displaymath}
  and $E_\eps$ is the remainder. As in the case $d>\m$, a contribution
  to $|\ln\eps|E_\eps$ comes from
  \begin{displaymath}
   \dint_0^t \dint_0^{\frac v{\eps^\m}} \dint
    e^{-\xi^\m v} (e^{\eps^\m\xi^\m s_1}-1) 
    e^{-\xi_1^\m s_1} \hat R(\xi_1-\eps\xi) d\xi_1 ds_1
   \Ues(t-v,\xi) dv. 
  \end{displaymath} 
  We again decompose the integral in $s_1$ into $0\leq s_1\leq
  \frac{v}{2\eps^\m}$ and $\frac{v}{2\eps^\m}\leq s_1\leq \frac
  v{\eps^\m}$. We have
  \begin{displaymath}
  \begin{array}{l}
   \dint_0^t \dint_{\frac v{2\eps^\m}}^{\frac v{\eps^\m}} \dint
    e^{-\xi^\m v} (e^{\eps^\m\xi^\m s_1}-1) 
    e^{-\xi_1^\m s_1} \hat R(\xi_1-\eps\xi) d\xi_1 ds_1 dv \\[4mm]
   \leq \dint \dfrac{2}{\eps^\m}\Big(\dfrac{\eps^\m}{\xi_1^\m}\wedge t\Big)^2
   \hat R(\xi_1-\eps\xi) d\xi_1 \lesssim \rho_f,
  \end{array}
  \end{displaymath}
  according to lemma \ref{lem:bound}. Also,
  \begin{displaymath}
  \begin{array}{l}
  \dint_0^t \dint_0^{\frac v{2\eps^\m}}\dint
    e^{-\xi^\m v} (e^{\eps^\m\xi^\m s_1}-1) 
    e^{-\xi_1^\m s_1} \hat R(\xi_1-\eps\xi) d\xi_1 ds_1 dv 
   \lesssim \eps^\m\big(\dfrac{t}{2\eps^\m} \vee 1\big)
  \end{array}
  \end{displaymath}
  according to the calculations performed above on $I_3$,
  which is uniformly bounded, and thus provides a $|\ln\eps|^{-1}$
  contribution to $E_\eps$.
  
  We are thus left with the analysis of
  \begin{displaymath}
   U(t,\xi)\mapsto \dint_0^t 
   e^{-\xi^\m v} \Big( \dfrac{1}{|\ln\eps|}
  \dint \dfrac{1-e^{-\frac{\xi_1^\m v}{\eps^\m}}}{\xi_1^\m}
  \hat R(\xi_1-\eps\xi) d\xi_1 -\rho_\eps\Big) U(t-v,\xi) dv,
 \end{displaymath}
 as an operator in $\mathcal L(\mathcal C(0,T))$ for $\xi$ fixed.
 Define $\hat R_\eps(\xi_1)=\hat R(\xi_1-\eps\xi)$. The integral in
 $\xi_1$ may be recast as
  \begin{displaymath}
  \dint_0^\infty \dfrac{1-e^{-\frac{r^\m v}{\eps^\m}}}{r}
   \Big(\dint_{S^{d-1}}\hat R_\eps(r\theta)d\mu(\theta)\Big) dr.
  \end{displaymath}
  We observe that the integral on $(1,\infty)$ is bounded by $\|\hat
  R\|_1$. Assuming that $\hat R$ is of class $\mathcal
  C^{0,\gamma}(\Rm^d)$ for $\gamma>0$, we write $\hat
  R_\eps(\xi_1)=\hat R_\eps(0)+(\hat R_\eps(\xi_1)-\hat R_\eps(0))$.
  The second contribution generates a term proportional to $r^\gamma$
  in the integral and thus is bounded independent of $\eps$. It
  remains to estimate
  \begin{displaymath}
   c_d \hat R_\eps(0)\dint_0^1 \dfrac{1-e^{-\frac{r^\m v}{\eps^\m}}}{r} dr
   =c_d\hat R_\eps(0) \dint_0^{\frac{v^{\frac1\m}}{\eps}}
   \dfrac{1-e^{-r^\m}}{r} dr.
  \end{displaymath}
  The latter integral restricted to $(0,1)$ is bounded. On $r\geq1$,
  $e^{-r^\m}/r$ is uniformly integrable so that
  \begin{displaymath}
    c_d\hat R_\eps(0)\dint_0^1 \dfrac{1-e^{-\frac{r^\m v}{\eps^\m}}}{r} dr=
    c_d\hat R(\eps\xi) |\ln\eps| + O(1).
  \end{displaymath}
  This shows that $E_\eps$ is of order
  $\frac{1}{|\ln\eps|}=\eps^\beta$ as an operator on $\mathcal C(0,T)$
  and concludes the proof of the lemma.
\end{proof}
Note that $A_\eps$ may be written as
\begin{displaymath}
  A_\eps U(t,\xi) = \dint_0^t \varphi_\eps(s,\xi) U(t-s\xi) ds,
\end{displaymath}
where $\varphi_\eps(s,\xi)$ is uniformly bounded in $s$, $\xi$, and
$\eps$ by a constant $\varphi_\infty$.  The equation
\begin{displaymath}
  (I-A_\eps) U(t,\xi) = S(t,\xi),
\end{displaymath}
admits a unique (by Gronwall's lemma) solution given by the Duhamel
expansion and bounded by 
\begin{displaymath}
  |U(t,\xi)| \leq \|S\|_\infty e^{t\varphi_\infty}.
\end{displaymath}

As in the proof of lemma \ref{lem:Eeps}, let us define
$B_\eps=A_\eps-E_\eps$.  We verify that $\UU_{\eps}(t,\xi)$, the
solution to
\begin{displaymath}
  (I-B_\eps)\UU_{\eps} = e^{-t\xi^\m}\hat u_\eps(\xi),
\end{displaymath}
is given by
\begin{equation}
  \label{eq:tildeUeps}
  \UU_{\eps}(t,\xi) = e^{-t(\xi^\m-\rho_\eps(\xi))} \hat u_0(\xi).
\end{equation}
The solution may thus grow exponentially in time for low frequencies.
The error $V_\eps(t,\xi)=(U_{\eps,s}(t,\xi)-\UU_{\eps}(t,\xi))$
is a solution to
\begin{displaymath}
  (I-A_\eps) V_\eps = E_\eps \UU_{\eps}(t,\xi),
\end{displaymath}
so that over bounded intervals in time (with a constant growing
exponentially with time but independent of $\xi$), we find that
\begin{equation}
  \label{eq:bdVeps}
  |V_\eps(t,\xi)| \lesssim \eps^\beta.
\end{equation}

Up to an order $O(\eps^\beta|\hat u_0(\xi)|)$, we have thus obtained
that $\E\{\hat u_\eps(t,\xi)\}$ is given by
\begin{displaymath}
  e^{-t(\xi^\m-\rho_\eps(\xi))} \hat u_0(\xi),
\end{displaymath}
which in the physical domain gives rise to a possibly non-local
equation.  It remains to analyze the limit of the above term, and thus
the error $\rho_\eps(\xi)-\rho$, which depends on the regularity of
$\hat R(\xi)$. For $\hat R(\xi)$ of class $\mathcal C^2(\Rm^d)$, we
find that
\begin{displaymath}
  \big|e^{-t(\xi^\m-\rho_\eps(\xi))}-e^{-t(\xi^\m-\rho)}\big| 
  \leq te^{Ct} e^{-\xi^\m t} \big|\rho_\eps(\xi)-\rho\big|
   \lesssim e^{Ct} e^{-\xi^\m t} \eps^2 t\xi^2.
\end{displaymath}
The reason for the second order accuracy is that $\hat R(-\xi)=\hat
R(\xi)$ and $\nabla \hat R(0)=0$ so that first-order terms in the
Taylor expansion vanish.  For $\hat R(\xi)$ of class $\mathcal
C^\gamma(\Rm^d)$ with $0<\gamma<2$, we obtain by interpolation that
\begin{displaymath}
  \big|e^{-t(\xi^\m-\rho_\eps(\xi))}-e^{-t(\xi^\m-\rho)}\big|
   \lesssim e^{Ct} e^{-\xi^\m t} \eps^\gamma t\xi^\gamma. 
\end{displaymath}
When $\m\geq\gamma$, the above term is bounded by $O(\eps^\gamma)$
uniformly in $\xi$ and uniformly in time on bounded intervals. When
$\m\leq\gamma$, the above term is bounded by $O(\eps^\m)$ uniformly in
$\xi$ and uniformly in time on bounded intervals. This concludes
the proof of theorem \ref{thm:convergence}. In terms of the propagators
defined in \eqref{eq:Uepsns}, we may recast the above result as
\begin{equation}
  \label{eq:errorprop}
   \big|{\mathcal U}_\eps(t,\xi)-{\mathcal U}(t,\xi)\big| \lesssim 
    \eps^{\gamma\wedge \beta},
  \qquad
   {\mathcal U}(t,\xi) = e^{-(\xi^\m-\rho)t},
\end{equation}
where the bound is uniform in time for $t\in (0,T)$ and uniform in
$\xi\in\Rm^d$.

\subsection{Fluctuation theory for $u_\eps$}

We now address the proof of theorem \ref{thm:fluct}.  The first term
in the decomposition of $\hat u_{n,\eps}$ defined in
\eqref{eq:hatuneps} is its mean $\E\{\hat u_{n,\eps}\}$, which was
analyzed in the preceding section.  The second contribution
corresponds to the graphs $\mG_{cs}$ in the analysis of the
correlation function and is constructed as follows. Let $n=2p+1$,
$p\in\Nm$. We introduce the corrector $\hat u_{n,\eps}^c$ given by
\begin{equation}
  \label{eq:hatunepsc}
  \begin{array}{l}
    \hat u_{n,\eps}^c(t,\xi_0) = \dint \dprod_{k=0}^n  e^{-s_k \xi_k^\m} 
    \dsum_{q=0}^p \Big[\dprod_{r=1}^{q}
   \E\{\hat q_\eps(\xi_{2(r-1)}-\xi_{2r-1})\hat q_\eps(\xi_{2r-1}-\xi_{2r})\}
   \Big] \\\qquad \qquad\hat q_\eps(\xi_{2q}-\xi_{2q+1})
    \Big[\dprod_{r=q+1}^{p}
   \E\{\hat q_\eps(\xi_{2r-1}-\xi_{2r})\hat q_\eps(\xi_{2r}-\xi_{2r+1})\}
   \Big]  \hat u_0(\xi_n) d\bs d\bxi.
  \end{array}
\end{equation}
In other words, all the random terms are averaged as simple pairs
except for one term. There are $p+1$ such graphs. We define
\begin{equation}
  \label{eq:hatuepsc}
  \hat u_\eps^c(t,\xi) = \dsum_{n\geq1} \hat u_{n,\eps}^c(t,\xi).
\end{equation}
We verify that 
\begin{displaymath}
 V_{\eps,s}^{n,m}(t,\xi_0,\xi_{n+m+1}) :=
  \E\{\hat u_{n,\eps}^c(t,\xi_0) \bar{\hat u}_{n,\eps}^c(t,\xi_{n+m+1})\}
\end{displaymath}
is equal to the sum in $V^{n,m}_\eps(t,\xi_0,\xi_{n+m+1})$ only over
the graphs in $\mG_{cs}$. Indeed, the above correlation involves all
the graphs composed of simple pairs with a single crossing.

Now let us define the variable
\begin{equation}
  \label{eq:Yeps}
  Y_\eps = (\hat u_\eps-\hat u_\eps^c-\E\{\hat u_\eps\},\hat M).
\end{equation}
Summing over $n,m\in\Nm$ the inequality in \eqref{eq:bounddeltaV} as
we did to obtain \eqref{eq:Xeps2}, we have demonstrated that
\begin{equation}
  \label{eq:bdYeps}
  \E\{Y_\eps^2\} \lesssim \eps^{d-2\alpha+\beta} \|\hat u_0\|^2
  \|\hat M\|_1^2,
\end{equation}
for sufficiently small times.  The leading term in the random
fluctuations of $u_\eps$ is thus given by $u_\eps^c$. It remains to
analyze the convergence properties of
\begin{equation}
  \label{eq:Zeps}
  Z_\eps(t) = \dfrac{1}{\eps^{\frac{d-2\alpha}2}} (\hat u_\eps^c,\hat M).
\end{equation}

We thus come back to the analysis of $\hat u_\eps^c$ and observe that
for $n=2p+1$,
\begin{displaymath}
  \begin{array}{l}
    \hat u_{n,\eps}^c(t,\xi_0) = \dint \dprod_{k=0}^n  e^{-s_k \xi_k^\m} 
    \dsum_{q=0}^p  
    \Big[\dprod_{r=1}^{q}
    \eps^{d-2\alpha} \hat R(\eps(\xi_{2r-1}-\xi_0))
    \delta(\xi_{2r}-\xi_0)
    \Big] \\\qquad \qquad\hat q_\eps(\xi_{0}-\xi_{n})
    \Big[\dprod_{r=q+1}^{p}
    \eps^{d-2\alpha} \hat R(\eps(\xi_{2r}-\xi_n))
    \delta(\xi_{2r-1}-\xi_n)
     \Big]  \hat u_0(\xi_n) d\bs d\bxi.
  \end{array}
\end{displaymath}
Using the propagator defined in \eqref{eq:Uepsns}, we verify that
\begin{displaymath}
  \begin{array}{l}
    \hat u_{n,\eps}^c(t,\xi_0) = \dsum_{q=0}^p  
    \dint \dprod_{k=0}^{2q+1}  e^{-s_k \xi_k^\m} 
    \Big[\dprod_{r=1}^{q}
    \eps^{d-2\alpha} \hat R(\eps(\xi_{2r-1}-\xi_0))
    \delta(\xi_{2r}-\xi_0)
    \Big] \\\qquad \qquad\hat q_\eps(\xi_{0}-\xi_{n})
    {\mathcal U}^{n-2q}_\eps(t_{2q+1},\xi_n) \hat u_0(\xi_n) d\tilde\bs 
    d\tilde\bxi\\
  = \dsum_{q=0}^p \dint_0^t {\mathcal U}^{2q}_\eps(t-t_{2q+1},\xi_0)
    \hat q_\eps(\xi_{0}-\xi_{n})
    {\mathcal U}^{n-2q}_\eps(t_{2q+1},\xi_n) \hat u_0(\xi_n)
    dt_{2q+1} d\xi_n\\
  = \dsum_{q=0}^p \dint_0^t {\mathcal U}^{2q}_\eps(t-s,\xi_0)
    \hat q_\eps(\xi_{0}-\xi_{1})
    {\mathcal U}^{n-2q}_\eps(s,\xi_1) \hat u_0(\xi_1)  ds d\xi_1.
  \end{array}
\end{displaymath}
Upon summing over $n$, we obtain
\begin{equation}
  \label{eq:uc}
  \hat u_\eps^c(t,\xi) = \dint_0^t {\mathcal U}_\eps(t-s,\xi)
    \hat q_\eps(\xi-\xi_1) {\mathcal U}_\eps(s,\xi_1) 
     \hat u_0(\xi_1)ds d\xi_1.
\end{equation}
We can use the error on the propagator obtained in
\eqref{eq:errorprop} to show that the leading order of $\hat u_\eps^c$
is not modified by replacing ${\mathcal U}_\eps$ by ${\mathcal U}$.
In other words, replacing ${\mathcal U}_\eps$ by ${\mathcal U}$
modifies $Z_\eps$ in \eqref{eq:Zeps} by a term of order
$O(\eps^{\frac12(\beta\wedge \gamma)})$ in $L^2(\Omega\times \Rm^d)$,
which thus goes to $0$ in law.

Note that $\hat u_\eps^c(t,\xi)$ is a mean zero Gaussian random
variable. It is therefore sufficient to analyze the convergence of its
variance in order to capture the convergent random variable for each
$t$ and $\xi$. The same is true for the random variable $Z_\eps$. Up
to a lower-order term, which does not modify the final convergence, we
thus have that
\begin{displaymath}
   (\hat u_\eps^c,\hat M) =\dint \dint_0^t 
   {\bar {\mathcal U}}_{\hat M}(t-s,\xi)
    \hat q_\eps(\xi_1) {\mathcal U}_{\hat u_0}(s,\xi-\xi_1) ds d\xi d\xi_1.
\end{displaymath}
We have defined ${\mathcal U}_f(t,\xi)={\mathcal U}(t,\xi) f(\xi)$ for
a function $f(\xi)$.  As a consequence, we find that, still up a
vanishing contribution,
\begin{displaymath}
   \begin{array}{rcl}
  \E\{|Z_\eps|^2\}& =&
   \dint \dint_0^t\dint_0^t {\bar{\mathcal U}}_{\hat M}(t-s,\xi)
   {\mathcal U}_{\hat M}(t-\tau,\zeta) \hat R(\eps\xi_1)
    \delta(\xi_1-\zeta_1) \\ && \qquad \qquad \times\,\,
    {\mathcal U}_{\hat u_0}(s,\xi-\xi_1)
   \bar{\mathcal U}_{\hat u_0}(\tau,\zeta-\zeta_1) 
  d[s\tau\zeta\zeta_1\xi\xi_1].
  \end{array}
\end{displaymath}
Here and below, we use the notation $d[x_1\ldots x_n]\equiv dx_1\ldots
dx_n$.  By the dominated Lebesgue convergence theorem, we obtain in
the limit
\begin{displaymath}
  \E\{|Z|^2\} := 
  \hat R(0) \dint \Big| \dint \dint_0^t{\mathcal U}_{\hat M}(t-s,\xi)
      {\mathcal U}_{\hat u_0}(s,\xi-\xi_1) d\xi_1 ds\Big|^2 d\xi.
\end{displaymath}
Here, $Z$ is defined as a mean zero Gaussian random variable with the 
above variance. Let us define
\begin{math}
  {\mathcal G}^\rho_t f(x),
\end{math}
the solution at time $t$ of \eqref{eq:parab} with $f(x)$ as
initial conditions, which is also the inverse Fourier transform
of ${\mathcal U}_{\hat f}(t,\xi)$. 
We then recognize in 
\begin{math}
  \int \int_0^t{\mathcal U}_{\hat M}(t-s,\xi)
      {\mathcal U}_{\hat u_0}(s,\xi-\xi_1) d\xi_1 ds
\end{math}
the Fourier transform of $\mathcal M_t(x)$ defined in \eqref{eq:conv}
so that by an application of the Plancherel identity, we find that
\begin{equation} \label{eq:varZ}
  \E\{Z^2\} = (2\pi)^d \hat R(0) \!\!\dint_{\Rm^d}
   \Big(\dint_0^t {\mathcal G}_{t-s}^\rho M(x) {\mathcal G}_s^\rho u_0(x) ds
   \Big)^2 dx = (2\pi)^d \hat R(0) \!\!\dint_{\Rm^d} \!\!\mathcal M_t^2(x)dx.
\end{equation}
This shows that $Z(t)$ is indeed the Gaussian random variable written
on the right hand side in \eqref{eq:conv} by an application of the
It\^o isometry formula. This concludes the proof of theorem
\ref{thm:fluct}.

\subsection{Long range correlations and correctors}
\label{sec:longrange}

Let us now assume that 
\begin{equation}
  \label{eq:decR}
  \hat R(\xi) = h(\xi) S(\xi),\qquad 0<h(\lambda\xi)=|\lambda|^{-\n}h(\xi),
\end{equation}
where $h(\xi)$ is thus a positive function homogeneous of degree $-\n$
and $\hat S(\xi)$ is bounded on $B(0,1)$. We assume that $\hat R(\xi)$
is still bounded on $\Rm^d\backslash B(0,1)$. We also assume that
$\m+\n<d$ and that $\rho$ in \eqref{eq:rho} is still defined.  We
denote by $\varphi(x)$ the inverse Fourier transform of $h(\xi)$.
Then we have the following result.
\begin{theorem}
  \label{thm:fracBM}
  Let us assume that $h(\xi)=|\xi|^{-\n}$ for $\n>0$ and $\m+\n<d$.
  We also impose the following regularity on $\hat u_0$:
  \begin{equation}
    \label{eq:regu0}
    \dint_{B(0,1)} |\hat u_0(\xi+\tau)|^2 h(\xi) d\xi \leq C,
   \quad \mbox{ for all }\, \tau\in\Rm^d.
  \end{equation}
  Then theorem \ref{thm:convergence} holds with $\beta$ replaced by
  $\beta-\n$. 

  Let us define the random corrector 
  \begin{equation}
  \label{eq:u1eps2}
    u_{1,\eps}(t,x) = \dfrac{1}{\eps^{\frac{d-\m-\n}2}}
    \big(u_\eps-\E\{u_\eps\}\big)(t,x).
  \end{equation}
  Then its spatial moments $(u_{1,\eps}(t,x),M(x))$ converge in law to
  centered Gaussian random variables $\mathcal N(0,\Sigma_M(t))$
  with variance given by
  \begin{equation}
    \label{eq:sigmaM}
    \Sigma_M(t) = (2\pi)^{d}\hat S(0) 
      \dint_{\Rm^{2d}} \mathcal M_t(x) \varphi(x-y) \mathcal M_t(y) dx dy.
  \end{equation}
\end{theorem}
\begin{proof}
  The proof of theorem \ref{thm:convergence} relies on three
  estimates: those of lemma \ref{lem:bound} and lemma \ref{lem:Eeps}
  and the uniform bound in \eqref{eq:unifbdhatR} for $\hat R$. Lemmas
  \ref{lem:bound} and \ref{lem:Eeps} were written to account for power
  spectra bounded by $|\xi|^{-\n}$ in the vicinity of the origin. It
  thus remains to replace \eqref{eq:unifbdhatR} by
  \begin{displaymath}
    \eps^{d-\m} \hat R(\eps(\xi_Q-\xi_{Q-1}^\eps))
   \leq \eps^{d-\m-\n} h(\xi_Q-\xi_{Q-1}^\eps) \hat S_\infty,
  \end{displaymath}
  when $|\xi_Q-\xi_{Q-1}^\eps|\leq1$ while we still use
  \eqref{eq:unifbdhatR} otherwise. We have defined $\hat S_\infty$ as
  the supremum of $\hat S(\xi)$ in $B(0,1)$. It now remains to show
  that the integration with respect to $\xi_Q$ in \eqref{eq:Vepsmnbd1}
  is still well-defined. Note that either $Q=n$ or
  $\xi_Q-\xi_{Q-1}^\eps$ may be written as $\xi_n-\zeta$ for some
  $\zeta\in\Rm^d$ thanks to \eqref{eq:deltause}.  Upon using
  \eqref{eq:ineq}, we thus observe that in all cases, the integration
  with respect to $\xi_Q$ in \eqref{eq:Vepsmnbd1} is well-defined and
  bounded uniformly provided that \eqref{eq:u1eps2} is satisfied
  uniformly in $\tau$. Using the H\"older inequality, we verify that
  \eqref{eq:u1eps2} holds e.g.  when $\hat u_0(\cdot-\tau)\in
  L^q(B(0,1))$ uniformly in $\tau$ for $q>\frac{2d}{d-\n}$. This
  concludes the proof of the first part of the theorem.

  Let us now define 
  \begin{displaymath}
    \tilde Z_\eps(t) = 
   \dfrac{1}{\eps^{\frac{d-\m-\n}2}} (\hat u_\eps^c,\hat M)
   = \eps^{\frac{\n}2} Z_\eps(t).
  \end{displaymath}
  We verify as for the derivation of $\E\{Z_\eps^2\}$ that 
  \begin{displaymath}
     \begin{array}{rcl}
  \E\{\tilde Z_\eps^2\}& =&
   \dint \dint_0^t\dint_0^t {\mathcal U}_{\hat M}(t-s,\xi)
   \hat{\mathcal U}_{\hat M}(t-\tau,\zeta)\hat S(\eps\xi_1)h(\xi_1)
    \delta(\xi_1-\zeta_1) \\ && \qquad \qquad \times\,\,
    {\mathcal U}_{\hat u_0}(s,\xi-\xi_1)
   \hat{\mathcal U}_{\hat u_0}(\tau,\zeta-\zeta_1) 
  d[s\tau\zeta\zeta_1\xi\xi_1].
     \end{array}
  \end{displaymath}
  The dominated Lebesgue convergence theorem yields in the limit
  $\eps\to0$
  \begin{displaymath}
     \begin{array}{rcl}
  \E\{\tilde Z^2\} &:=& 
  \hat S(0) \dint \Big|\dint_0^t \dint {\mathcal U}_{\hat M}(t-s,\xi)
      {\mathcal U}_{\hat u_0}(s,\xi-\xi_1) h^{\frac12}(\xi_1)
    d\xi_1 ds\Big|^2 d\xi \\
   &=& \hat S(0) \dint |\hat{\mathcal M}_t (\xi)|^2  h(\xi)  d\xi,
     \end{array}
  \end{displaymath}
  where ${\mathcal M}_t$ is defined in \eqref{eq:conv}.  An
  application of the inverse Fourier transform yields
  \eqref{eq:sigmaM}.
\end{proof}
Note that \eqref{eq:sigmaM} generalizes \eqref{eq:varZ}, where
$\varphi(x)=\delta(x)$, to functions $\mathcal M_t(x)\in
L^2_\varphi(\Rm^d)$  with inner product
\begin{equation} \label{eq:innerprod}
  (f,g)_\varphi = \dint_{\Rm^{2d}} f(x)g(y)\varphi(x-y) dxdy.
\end{equation}
For $h(\xi)=|\xi|^{-\n}$, we find that $\varphi(x)=c_{\n}|x|^{\n-d}$,
with $c_{\n}=\Gamma(\frac{d-\n}2)/(2^\n\pi^{\frac
  d2}\Gamma(\frac{\n}2))$ a normalizing constant. Following e.g.
\cite{HOUZ-Birk-96,L-BLM-93}, we may then define a stochastic integral
with fractional Brownian
\begin{equation} \label{eq:Z}
   Z = \dint_{\Rm^d} {\mathcal M}_t (x) dB^H(x),
\end{equation}
where $B^H$ is fractional Brownian motion such that 
\begin{displaymath}
  \E\{ B^H(x)B^H(y) \} = \frac12\big( |x|^{2H}+|y|^{2H}-|x-y|^{2H}\big), 
   \qquad 2H = 1+\frac{\n}d.
\end{displaymath}
We then verify that $\E\{Z^2\}=\Sigma_M$ so that the random variable
$Z$ is indeed given by the above formula \eqref{eq:Z}.  When $\n=0$,
we retrieve the value for the Hurst parameter $H=\frac 12$ so that
$B^H=W$.  The above isotropic fractional Brownian motion is often
replaced in the analysis of stochastic equations by a more Cartesian
friendly fractional Brownian motion defined by
\begin{displaymath}
  \varphi_H(x) = \dprod_{i=1}^d H_i (2H_i-1)|x_i|^{2H_i-2}.
\end{displaymath}
The above is then defined as the Fourier transform of 
\begin{displaymath}
  h_H(\xi) = \dprod_{i=1}^d  |\xi_i|^{-\n_i}, \qquad
   \dsum_{i=1}^d \n_i = \n, \qquad 2H_i = 1 + \dfrac{\n_i}d.
\end{displaymath}
The results of theorem \ref{thm:convergence} and \ref{thm:fracBM} may
also be extended to this framework by modifying the proofs in lemmas
\ref{lem:bound} and \ref{lem:Eeps}. We then obtain that \eqref{eq:Z}
holds for a multiparameter anisotropic fractional Brownian motion
$B^H$, $H=(H_1,\ldots,H_d)$, with covariance
\begin{displaymath}
  \E\{ B^H(x)B^H(y) \}= \frac{1}{2^d}\prod_{i=1}^d 
   \big(|x_i|^{2H_i}+|y_i|^{2H_i}-|x_i-y_i|^{2H_i}\big).
\end{displaymath}
Note that homogenization theory is valid as soon as $d>\m+\n$. As in
the case $\n=0$, we expect that when $d<\m+\n$ (rather than $d<\m$),
the limit for $u_\eps$ will be the solutions in $L^2(\Omega\times
\Rm^d)$ to a stochastic differential equation of the form
\eqref{eq:stoch} with white noise replaced by some fractional Brownian
motion; see also \cite{Hu-AMO-01}.

The stochastic representation in \eqref{eq:Z} is not necessary since
$\Sigma_M(t)$ fully characterizes the random variable $Z$. However,
the representation emphasizes the following conclusion.  Let $Z_1^H$
and $Z_2^H$ be the limiting random variables corresponding to two
moments with weights $M_1(x)$ and $M_2(x)$ and a given Hurst parameter
$H$. When $H=\frac12$, we deduce directly from \eqref{eq:Z} that
$\E\{Z_1^{\frac12}Z_2^{\frac12}\}=0$ when $M_1(x)M_2(x)=0$, i.e., when
the supports of the moments are disjoint. This is not the case when
$H\not=\frac12$ as fractional Brownian motion does not have
independent increments. Rather, we find that $\E\{Z_1^{H}Z_2^{H}\}$ is
given by $(\mathcal M_{t,1},\mathcal M_{t,2})_{\varphi}$, where the
inner product is defined in \eqref{eq:innerprod} and $\mathcal
M_{t,k}$ is defined in \eqref{eq:conv} with $M$ replaced by $M_k$,
$k=1,2$. Similar results were obtained in the context of the
one-dimensional homogenization with long-range diffusion coefficients
\cite{BGMP-AA-08}.

\section*{Acknowledgment}

This work was supported in part by NSF Grants DMS-0239097 and
DMS-0804696.

\bibliography{../../bibliography} \bibliographystyle{siam}

\begin{thebibliography}{10}

\bibitem{B-CLH-08}
{\sc G.~Bal}, {\em Central limits and homogenization in random media},
  Multiscale Model. Simul., 7(2) (2008), pp.~677--702.

\bibitem{B-CMP-2-08}
\leavevmode\vrule height 2pt depth -1.6pt width 23pt, {\em {Convergence to
  SPDEs in Stratonovich form}}, submitted,  (2008).

\bibitem{BGMP-AA-08}
{\sc G.~Bal, J.~Garnier, S.~Motsch, and V.~Perrier}, {\em Random integrals and
  correctors in homogenization}, Asymptot. Anal.,  (2008).

\bibitem{Chen-JSP-05}
{\sc T.~Chen}, {\em {Localization lengths and Boltzmann limit for the Anderson
  model at small disorders in dimension 3}}, J. Stat. Phys.,  (2005),
  pp.~279--337.

\bibitem{Erdos-Yau2}
{\sc L.~Erd{\"o}s and H.~T. Yau}, {\em {Linear Boltzmann equation as the weak
  coupling limit of a random Schr\"odinger Equation}}, Comm. Pure Appl. Math.,
  53(6) (2000), pp.~667--735.

\bibitem{FOP-SIAP-82}
{\sc R.~Figari, E.~Orlandi, and G.~Papanicolaou}, {\em {Mean field and Gaussian
  approximation for partial differential equations with random coefficients}},
  SIAM J. Appl. Math., 42 (1982), pp.~1069--1077.

\bibitem{HOUZ-Birk-96}
{\sc H.~Holden, B.~{\O}ksendal, J.~Ub{\o}e, and T.~Zhang}, {\em Stochastic
  partial differential equations. A modeling, white noise functional
  approach.}, Probability and its applications, Birkh\"auser, Boston, MA, 1996.

\bibitem{Hu-AMO-01}
{\sc Y.~Hu}, {\em Heat equations with fractional white noise potentials}, Appl.
  Math. Optim., 43 (2001), pp.~221--243.

\bibitem{Hu-PA-02}
\leavevmode\vrule height 2pt depth -1.6pt width 23pt, {\em Chaos expansion of
  heat equations with white noise potentials}, Potential Anal., 16 (2002),
  pp.~45--66.

\bibitem{L-BLM-93}
{\sc T.~Lindstr{\o}m}, {\em {Fractional Brownian fields as integrals of white
  noise}}, Bull. London Math. Soc., 25 (1993), pp.~83--88.

\bibitem{LS-ARMA-07}
{\sc J.~Lukkarinen and H.~Spohn}, {\em Kinetic limit for wave propagation in a
  random medium}, Arch. Ration. Mech. Anal., 183 (2007), pp.~93--162.

\bibitem{NR-JFA-97}
{\sc D.~Nualart and B.~Rozovskii}, {\em Weighted stochastic sobolev spaces and
  bilinear spdes driven by space-time white noise}, J. Funct. Anal., 149
  (1997), pp.~200--225.

\bibitem{PP-GAK-06}
{\sc E.~Pardoux and A.~Piatnitski}, {\em {Homogenization of a singular random
  one dimensional PDE}}, GAKUTO Internat. Ser. Math. Sci. Appl., 24 (2006),
  pp.~291--303.

\end{thebibliography}

\end{document}